\documentclass{article}

\usepackage{arxiv}
\usepackage[utf8]{inputenc} 
\usepackage[T1]{fontenc}    
\usepackage{hyperref}       
\usepackage{url}            
\usepackage{booktabs}       
\usepackage{amsfonts}       
\usepackage{nicefrac}       
\usepackage{microtype}      
\usepackage{graphicx}
\usepackage{subcaption}
\usepackage{natbib}
\usepackage{doi}
\usepackage{amsmath}
\usepackage{multirow}
\usepackage{array}
\usepackage{booktabs}  
\usepackage{algorithm}
\usepackage{algorithmicx}
\usepackage{algpseudocode}
\usepackage{soul}
\title{tipping point sensitivity analysis for missing data in time-to-event endpoints: model-based and ad hoc approaches}

\author{ \href{https://orcid.org/0000-0001-8143-5011}{\includegraphics[scale=0.06]{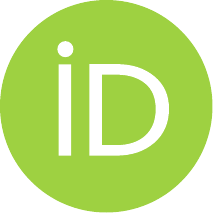}\hspace{1mm}Ajmal Oodally}\thanks{Corresponding author: ajmal.oodally@novartis.com}\hspace{2mm}\thanks{Equal contribution}\\
	Advanced Quantitative Sciences\\
	Novartis\\ 
     \\
	\And
	\href{https://orcid.org/0000-0003-1804-2463}{\includegraphics[scale=0.06]{orcid.pdf}\hspace{1mm}Craig Wang}\footnotemark[2] \\
	Advanced Quantitative Sciences\\
	Novartis\\
	\And
	\href{https://orcid.org/0000-0002-7655-0290}{\includegraphics[scale=0.06]{orcid.pdf}\hspace{1mm}Zheng Li} \\
	Advanced Quantitative Sciences\\
	Novartis\\ 
    \And
    \href{https://orcid.org/0000-0001-5850-3610}{\includegraphics[scale=0.06]{orcid.pdf}\hspace{1mm}Tim Morris} \\
	Advanced Quantitative Sciences\\
	Novartis Pharma AG\\ 
    \And
    \href{https://orcid.org/0000-0002-4111-1941}{\includegraphics[scale=0.06]{orcid.pdf}\hspace{1mm}Tobias M\"utze} \\
	Advanced Quantitative Sciences\\
	Novartis Pharma AG\\ 
    \And
	\href{https://orcid.org/0009-0008-9253-6414}{\includegraphics[scale=0.06]{orcid.pdf}\hspace{1mm}Arunava Chakravartty} \\
	Global Statistical Sciences\\
    Eli Lilly and Company\\ 
}

\renewcommand{\shorttitle}{Model-based and ad hoc tipping point analysis}
\newcommand{\boxedtext}[1]{\ifx#1\@empty\else\fbox{\parbox{.97\textwidth}{\textbf{Key Points}\par#1\vspace{-8pt}}}\fi}

\hypersetup{
  pdftitle={Model-based and ad hoc tipping point analysis},
  pdfsubject={Tipping Point Analysis},
  pdfauthor={Novartis}
}

\begin{document}
\maketitle

\begin{abstract}
Treatment policy estimands are frequently favored by regulators, as they assess the effect of treatment assignment regardless of post-randomization events. Despite best efforts, missing data due to study discontinuation cannot be fully avoided and, for time-to-event endpoints, typically manifests as right censoring. Study discontinuation is often more likely following intercurrent events, particularly when it coincides with treatment discontinuation, raising concerns about violations of the independent censoring assumption. Although the independent censoring assumption is routinely adopted for the main analyses, it may be unrealistic in practice and could lead to biased estimation of the treatment effect under the treatment policy estimand. Tipping-point analyses provide a structured framework to assess the robustness of trial conclusions to departures from the independent censoring assumption. This paper describes and contrasts model-based and two ad~hoc tipping point approaches, which involve “landmark” or “percentile sampling” based imputation. We illustrate their application using re-constructed examples based on real clinical trials, highlighting their underlying assumptions and implications for interpretation and clinical plausibility assessments of different tipping point approaches.
\end{abstract}

\section{Introduction}

The treatment policy estimand has emerged as one of the most commonly targeted estimands in confirmatory clinical trials and is frequently favored by regulatory authorities. Under this strategy, the effect of a decision to initiate treatment is evaluated regardless of post-randomization intercurrent events, such as treatment discontinuation or the initiation of subsequent therapies \citep{ICH_E9_R1_2019}. This estimand often aligns with intention-to-treat principles and in time-to-event settings is implicitly targeted by standard survival analysis methods, such as the Kaplan–Meier estimator and the Cox proportional hazards model, provided the underlying assumptions required for their validity hold.

Despite careful trial design and conduct, missing data cannot be fully prevented. Common causes include withdrawal of consent, and loss to follow-up. Although strategies such as enhanced retention efforts and systematic follow-up procedures can reduce their frequency, some degree of study discontinuation is inevitable in practice \citep{little2012prevention}. For time-to-event endpoints, missingness typically manifests as right censoring, most often due to patient discontinuing from study or administrative study termination. 
Importantly, study discontinuation is often related to the occurrence of intercurrent events. For example, discontinuation of study participation may coincide with discontinuation of active treatment due to adverse events, perceived lack of efficacy, or patient preference. In oncology and other open-label settings, discontinuation patterns may differ between treatment arms, reflecting underlying differences in tolerability, prognosis, or patient expectations. When discontinuation is associated with unobserved disease progression or prognosis, censoring may depend on unobserved event times even after conditioning on observed covariates. In such settings, the assumption that censoring is independent of the event process may be violated.

Within the missing data framework, mechanisms are commonly classified as missing completely at random (MCAR), missing at random (MAR), or missing not at random (MNAR), depending on the relationship between the probability of missingness and observed or unobserved data \citep{rubin1976inference}. For time-to-event outcomes, a direct analogue exists: censoring completely at random, censoring at random (CAR), and censoring not at random (CNAR). By convention, standard survival analyses assume CAR, meaning that conditional on covariates, the event time process is independent of the censoring time process \citep{Atkinson2019}. This assumption does not require censoring to be unrelated to treatment or prognosis per se, but rather that any such dependence can be removed by conditioning on observed covariates.

The independent censoring (CAR) assumption is often adopted as a default choice because of its interpretability and computational simplicity. Under CAR, likelihood-based survival estimators such as the Kaplan–Meier estimator and the Cox proportional hazards model consistently estimate the survival distribution or hazard ratio corresponding to the treatment policy estimand, provided the analysis model is correctly specified. However, in many practical settings the CAR assumption may be unrealistic. When censoring depends on unobserved outcomes even after conditioning on observed covariates, meaning that censoring is not at random (CNAR), the censoring mechanism is commonly referred to as informative censoring. In this case, standard survival analyses no longer identify the treatment policy estimand without additional, unverifiable assumptions. Treating such censoring as CAR may lead to biased estimation of treatment effects under a treatment policy estimand.

Regulatory guidance from agencies such as the US FDA and EMA, including the principles outlined in ICH E9 (R1) \citep{ICH_E9_R1_2019}, emphasizes the importance of sensitivity analyses to assess the robustness of conclusions to departures from the assumptions required for the primary analysis of time-to-event endpoints. Various approaches may be used to explore robustness, including ad hoc methods that regulators often request in settings with asymmetric patient discontinuation from study. We emphasize that asymmetry in study discontinuation between treatment arms is not inherently problematic, as it does not in itself imply informative censoring. However, in clinical trials, particularly randomized trials, the asymmetry may reflect systematic differences between patients who discontinue and those who remain under follow-up, for example with respect to prognosis or underlying disease status. In such situations, the censoring mechanism may become informative.

Tipping point analyses provide a structured and transparent framework for quantifying how strongly assumptions about the censoring mechanism must be altered to change (‘tip’) the study conclusions. Rather than asserting the plausibility of any particular assumption, these analyses explore the range of departures from CAR under which the primary results would no longer hold. The extent of deviation required to overturn a positive result offers a descriptive measure of robustness, although it does not validate any specific censoring assumption. Tipping point analyses have been widely developed for continuous and binary endpoints \citep{yan2009missing, liublinska2014sensitivity, gorst2022fast, sui2023application}, have been increasingly discussed in regulatory contexts \citep{Lipkovich2016, Fallah2024}, and have methodological extensions to time-to-event data \citep{Atkinson2019, Lipkovich2016}. However, practical guidance linking tipping point constructions to the primary estimand and analysis model in survival settings remains limited.

The aim of this paper is not to introduce new methodology, but to summarize commonly used tipping point analysis approaches in practice and provide guidance on their construction and interpretation for time-to-event endpoints, explicitly linking them to the primary analysis model and estimand of interest. Using two motivating clinical trial examples in which independent censoring is questionable, we illustrate different tipping point implementations and clarify the assumptions they implicitly encode. We discuss the degree to which particular constructions align with the treatment policy estimand and the primary survival model, and conclude with practical recommendations for their use in regulatory clinical trial settings, together with an R package implementation of the methods discussed \citep{tipse}.

\section{Motivating examples}
\label{sec:examples}
In the following subsections, we examine two clinical trials using their re-constructed data based on digitized Kaplan–Meier plots published in the literature, as well as two simulated examples with an imbalance in study discontinuation rates and timing of discontinuation between the experimental and control arms. While imbalance in censoring patterns alone is not a problem, if imbalance is also modified by prognosis, it may lead to bias. In the two clinical trials, the imbalances observed in discontinuation rates are suspected to be related to patients' prognosis. In the two simulated examples, we purposely introduce prognostic factors influencing the event and censoring process. These examples illustrate a common challenge in clinical research, where the imbalance in study discontinuation rates may raise concerns about the validity of trial results and the robustness of treatment effects to the assumption. These examples highlight the importance of sensitivity checks such as tipping point analysis, which can offer valuable insights into how different study discontinuation scenarios might influence results. However, despite the relevance of such analyses, they are often not explored or inconsistently used in clinical trials. This paper therefore advocates for more thorough and systematic tipping-point evaluations in order to ensure the reliability and accuracy of conclusions drawn from clinical studies.

\subsection{CodebreaK200}
\begin{figure}[ht]
    \centering
    \includegraphics[width=0.75\linewidth]{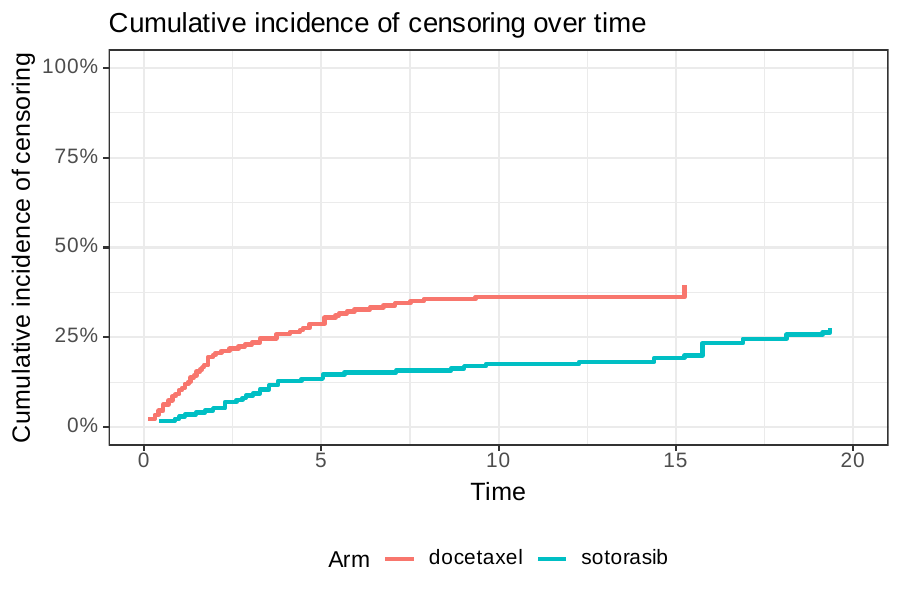}
    \caption{Cumulative incidence rate for censoring due to early study discontinuation in the CodebreaK200 trial}
    \label{fig:km_ttc1}
\end{figure}

The randomized, open-label phase 3 trial enrolled patients with advanced non–small cell lung cancer (NSCLC) harboring a KRASG12C mutation, randomly assigning them in a 1:1 ratio to receive either experimental treatment sotorasib or control treatment docetaxel. CodeBreaK200 demonstrated a statistically significant but modest improvement in the primary endpoint PFS (hazard ratio [HR], 0.66; 95$\%$ confidence interval [CI], 0.51–0.86; one-sided P = .002) as presented in \cite{de2023sotorasib}. The disproportionate early study discontinuation rates between the two arms is notable, with a higher proportion of patients in the control arm (13$\%$) never receiving the assigned study drug compared to the sotorasib arm (1.2$\%$). The disparity in early study discontinuation between the two arms is evident in cumulative incidence curves presented in Figure \ref{fig:km_ttc1} (a), particularly in the sharp increase observed at the start of the trial. Most of the untreated patients in the control arm withdrew consent within five weeks of randomization and were censored for progression-free survival (PFS) on day 1 due to the absence of post-baseline imaging in the trial. 

\subsection{ExteNET}
\begin{figure}[ht]
    \centering
    \includegraphics[width=0.75\linewidth]{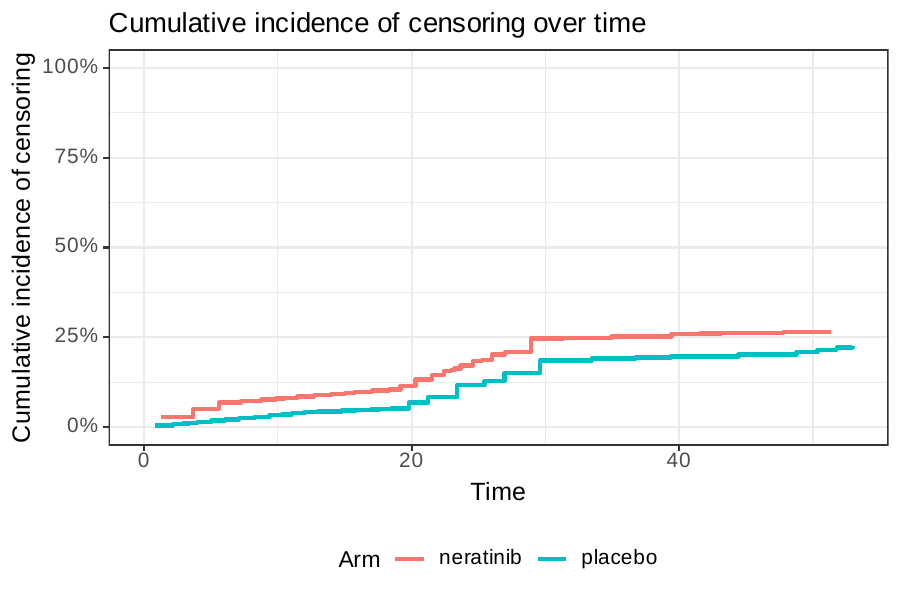}
    \caption{Cumulative incidence rate for censoring due to early study discontinuation in the ExteNET trial}
    \label{fig:km_ttc2}
\end{figure}

The ExteNET study was a randomized, double-blind, placebo-controlled, phase 3 trial that evaluated neratinib as an extended adjuvant therapy in patients with early-stage HER2-positive breast cancer who had completed trastuzumab-based treatment. The trial demonstrated that neratinib significantly improved invasive disease-free survival (iDFS) compared to placebo. The primary analysis showed that neratinib improved 2-year iDFS versus placebo (stratified HR = 0.66, 95\% CI 0.49–0.90; P = 0.008) as presented in \cite{martin2017neratinib}. However, there was a notable imbalance in early study discontinuation rates between the two arms, with a higher discontinuation rate in the neratinib group (5.6\%) compared to the control group (1.8\%) . This imbalance raised concerns about potential informative censoring leading to bias in the results, as it could impact the robustness of the efficacy findings. The difference in early study discontinuation between the two arms is illustrated in the cumulative incidence curves shown in Figure \ref{fig:km_ttc2}, with a notable increase for neratinib at the beginning of the trial.

\subsection{Simulated examples}
\begin{figure}[ht]
    \centering
    \includegraphics[width=0.49\linewidth]{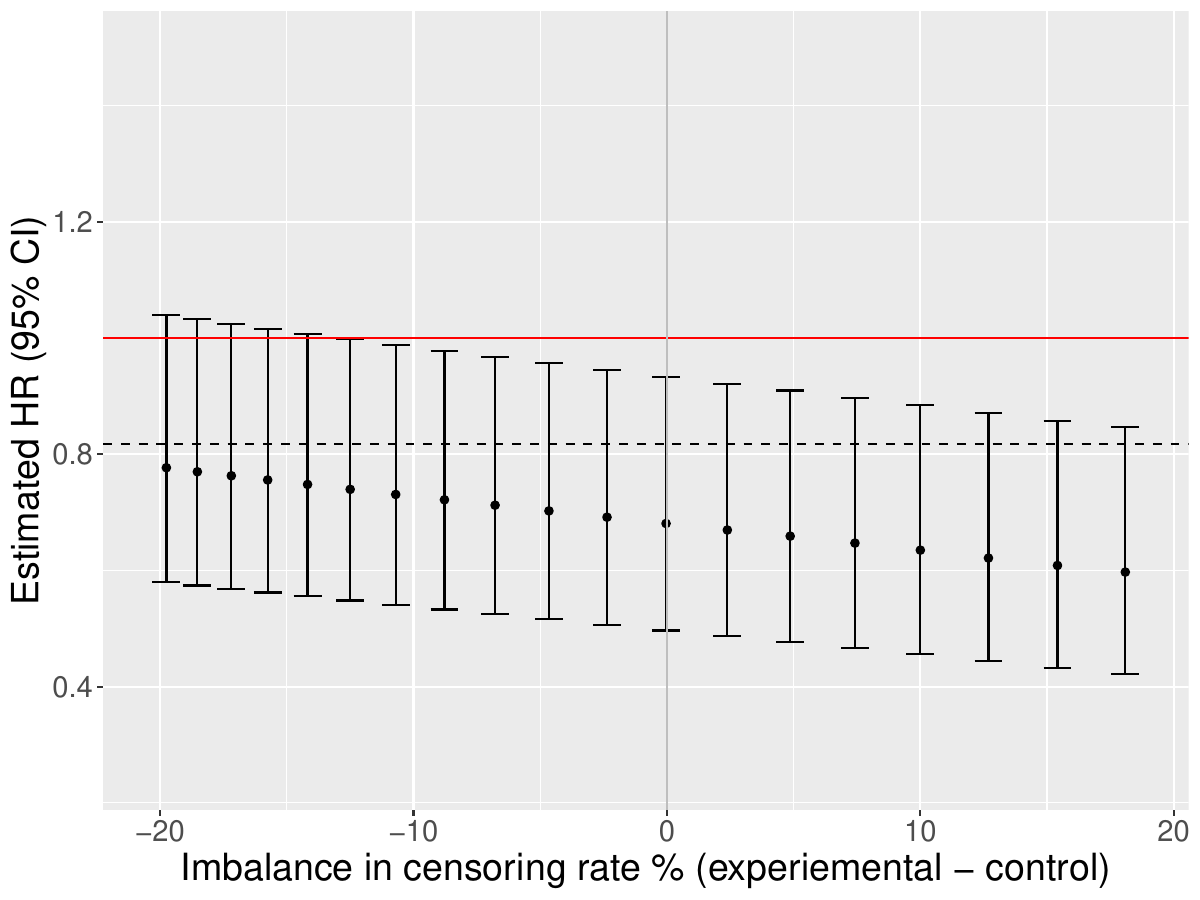}
    \includegraphics[width=0.49\linewidth]{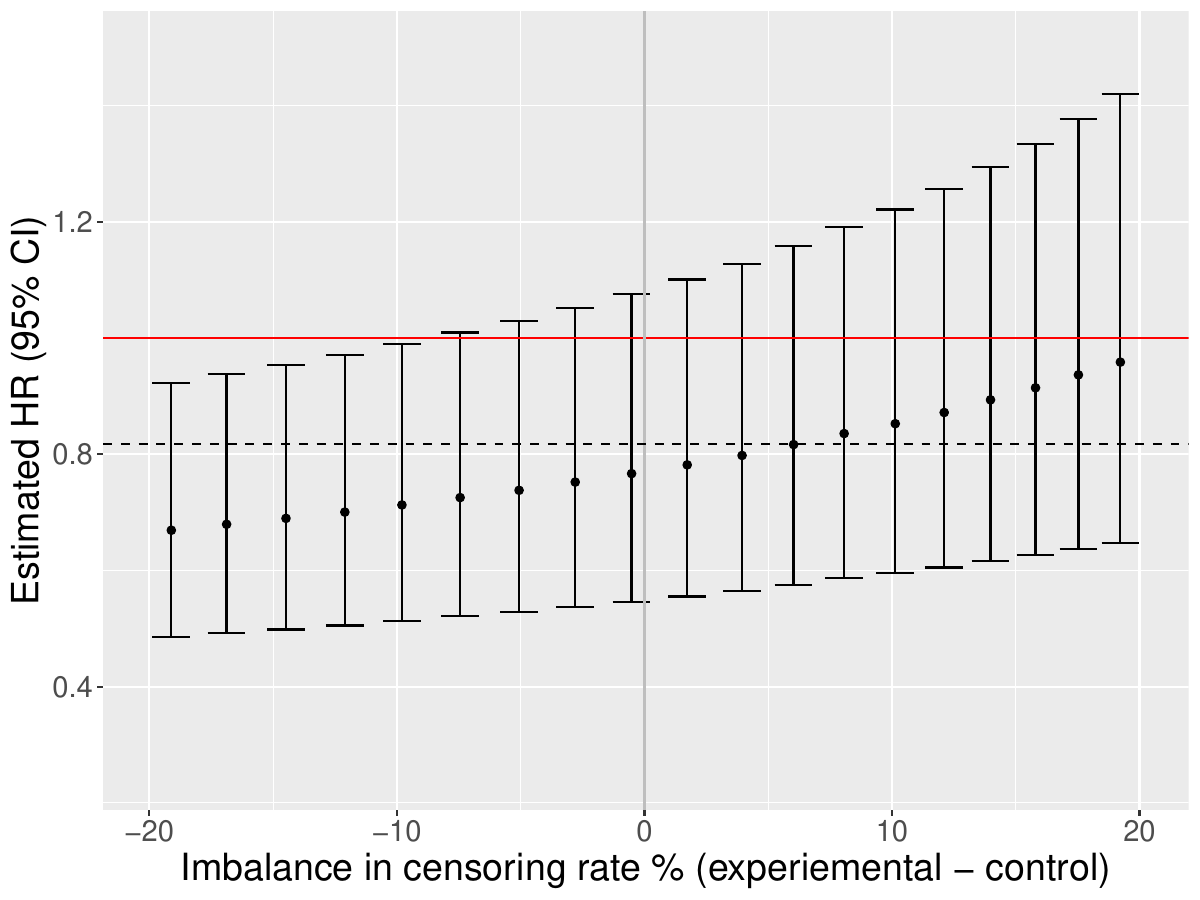}
    \caption{Simulated data with true marginal HR shown in horizontal dashed lines. The scenarios are simulated under different imbalances of informative censoring rate driven by prognostic factors. Under non-informative censoring, the HRs are expected to remain stable despite imbalanced censoring rate.}
    \label{fig:prob}
\end{figure}

Two simulated scenarios are created for illustrative purposes. in Figure~\ref{fig:prob}
\begin{itemize}
    \item[a)] higher censoring rate in experimental arm patients with worse prognosis.
    \item [b)] higher censoring rate in control arm patients with better prognosis.
\end{itemize}
It is intuitively clear in both cases that the interaction between randomized arm and prognosis could lead to overestimation of the difference between arms. Details of the data-generating are described in \hyperref[sec:sim]{Appendix}. 

Scenario a) refers to trials where patients with worse prognosis are more likely to discontinue from study. This represents a trial where patients in the experimental arm discontinue due to their inability to tolerate the drug. Such propensity to study discontinuation may be further caused by underlying prognostic factors. These early study discontinuations in the experimental arm may lead to an underestimation of hazard in the experimental arm and bias the trial outcome in its favor under treatment policy estimand. Scenario b) represents a possible open label study where patients with better prognosis on the control arm are more likely to discontinue from study. Such scenario can occur in trials where access to other therapies is available, leading to low-prognostic-risk patients in the control arm to discontinue. This imbalance may lead to an outcome that favors the experimental arm by overstating the hazard in the control arm. Beyond these illustrative scenarios, informative censoring may still arise despite similar overall censoring rates across arms when censoring is driven by prognostic covariates, thereby leading to biased treatment effect estimates.

\section{Methods for imputation and tipping point analysis}
This section introduces notation and describes imputation methods underlying three tipping-point approaches, two ad hoc and one model-based. The “model-based” approach is exactly as its name implies: a statistical model is used and dependent censoring mechanisms are injected by changing aspects of this model. The term “ad~hoc”, on the other hand are not based on a model. This should not be taken to imply free from modelling assumptions. Rather, these methods are termed “ad~hoc” in the missing data literature \citep{schafer_graham_2002, sterne_2009} in the sense that they are based on heuristic arguments and lack theoretical justification. In this context, there is no guarantee that embedding these approaches in a multiple imputation procedure will lead to reasonable tipping-point inference \citep{Meng1994}. They are, nevertheless, often used in regulatory contexts for their computational simplicity \citep{FDA2023_codebreak200}.

Denote by $T$ the time-to-event random variable with the survival function defined by $S(t) = P(T > t)$ and the hazard function $h(t)$ as the event rate at time $t$ conditional on survival until time $t$. 
Without loss of generality, we assume a two-arm trial under simple randomization with experimental treatment ($A_i = 1$) and control ($A_i = 0$) for $i = 1,...,N$. 

Tipping point analysis relies on a multiple imputation mechanism which can accommodate one or more varying sensitivity parameters to impute the missing data, enabling us to see the sensitivity parameter values under which a favorable treatment effect would be lost – the tipping point – based on either the point estimate or confidence interval crossing the null. Based on some selection criteria such as censoring reasons and experimental arm assignment, the event or censoring times from a subset of censored patients with size $N_c$ will be imputed. We describe the general model-based and ad hoc approaches and how they relate to each other, with a comparison summarized in Table \ref{tab:compare}.

\begin{table}[ht]
\centering
\renewcommand{\arraystretch}{1.3}
\begin{tabular}{|p{3cm}|p{4cm}|p{4cm}|p{4cm}|}
\hline
\textbf{Steps} & \textbf{Model-based} & \textbf{Percentile sampling} & \textbf{Landmark Sampling} \\
\hline

\textbf{1. Identify pool of candidate patients for imputation} & 
\multicolumn{3}{|p{12cm}|}{Identify \(N_C\) patients who were censored due to reasons suspected of informative censoring, making them candidates for imputation (typically based on randomized arm and reason for censoring, e.g. early discontinuation due to adverse events).} \\ \hline

\textbf{2. Define sensitivity parameter range} & 

\((\delta_1,\ldots,\delta_J)\) specifies the set of multiplicative factors \(\delta_j\) that will be used to modify the parametric hazard function fitted assuming independent censoring.

For \textbf{treatment arm}, the hazard will be inflated after their censoring time. 

For \textbf{control arm}, the hazard will be deflated after their censoring time. & 

\((\theta_1,\ldots,\theta_J)\) specifies the centiles \(\theta_j\) of the observed survival times from which will be sampled to impute censored patients.

For \textbf{treatment arm}, use the worst percentiles (shortest survival times) from the observed data of both arms. 

For \textbf{control arm}, use the best percentiles (longest survival times) from the observed data of both arms. & 

\((\kappa_1,\ldots,\kappa_J)\) specifies the number of patients \(\kappa_j\) whose event times will be imputed from the total number \(N_c\) of identified patients. 

For \textbf{treatment arm}, those patients will be imputed as having an event at the same timepoint when they were censored.

For \textbf{control arm}, those patients will be imputed as event-free at the landmark of study data cut-off. \\ \hline

\textbf{3. Impute for each sensitivity parameter value} & 
For \(\delta_j \in (\delta_1,\ldots,\delta_J)\), impute all candidate patients’ survival times after censoring using the parametric hazard function modified by \(\delta_j\), conditional on their observed censoring times & 
For \(\theta_j \in (\theta_1,\ldots,\theta_J)\), impute all candidate patients’ survival times by a random sample from the pool of \(\theta_j\%\) best/worst observed survival times. & 
For \(\kappa_j \in (\kappa_1,\ldots,\kappa_J)\), randomly select a subset of \(\kappa_j\) patients from the $N_c$ censored patients to impute as event time at their time of censoring, or at the data cut-off time. Those patients not selected do not have survival times imputed. \\
\hline

\end{tabular}
\caption{Comparison of different tipping point analysis approaches. In all cases, \(T_i^*\geq C_i\). Following imputation, inference is via Rubin’s rules and results are compared across the sensitivity parameter range.}
\label{tab:compare}
\end{table}

\subsection{Model-based imputation}
In model-based imputation, a specific choice of parametric model \citep{kleinbaum2012parametric} with parameters $\beta$ is fitted separately to each arm using all available data. Based on the fitted models and the intended imputation, a sensitivity parameter $\delta$ is introduced to modify the estimated hazard or the survival function after the censoring time $c_i$ as
$$h_A'(t;\beta|t>c_i) = \delta \hat{h}_A(t;\beta); S_A'(t|t>c_i) = \hat{S}_A(t;\beta)^\delta, \hspace{4mm}0< \delta < \infty $$ 

The corresponding estimated survival function is defined as
\[
\hat S_A(t;\beta)
= \exp\!\left(-\int_0^t \hat h_A(u;\beta)\, \mathrm{d}u \right).
\]

In model-based imputation, this model is fitted assuming independent censoring, and then the hazard function modified as above to simulate event times for censored individuals requiring imputation. This is embedded in a multiple imputation procedure and so uses proper multiple imputation that accounts for parameter uncertainty before simulating survival times. This is also known as the delta-adjustment method, thoroughly described in \citep{Lipkovich2016} and \citep{Jackson2014}.

Depending on the choice of $\delta$, it is possible for the imputed event times to exceed the maximum follow-up time on the study for some patients. Such inconsistences are avoided by introducing a truncation step such that any imputed event time that exceeds this threshold is censored to the maximum follow-up time. This mimics the process of administrative censoring in many time-to-event trials. For some scenarios, there may be patients whose potential follow-up times are longer than the observed maximum trial follow-up time. This leads to a limitation when imputing event times based on the Cox model and Kaplan–Meier methods, which cannot extrapolate the hazard or survival function beyond observed time windows hence they were not discussed in this paper. The parametric approaches are commonly used in practice for model-based imputation rather than semi- or non-parametric approaches. This is primarily due to their simplicity in taking draws from the posterior predictive distribution, but two further advantages are the flexibility of different parametric formulations and ability to extrapolate beyond observed time windows. See Algorithm~\ref{algo:model_based} for a step-by-step implementation of model-based tipping point analysis. 

\begin{algorithm}
\caption{Model-based imputation}
\label{algo:model_based}
\begin{algorithmic}[1]
    \State \textbf{Input:} survival data $(T, C, A)$ where $T$ and $C$ are event and censoring time, and $A$ is the arm assignment.
    \State \textbf{Identify:} early censored patients requiring imputation, either in the experimental or the control arm
    \State \textbf{Set:} appropriate ranges of hazard inflation factors ($1 \leq \delta_1, ..., \delta_J$) or deflation factors ($0\leq\delta_1, ..., \delta_J < 1)$
    \State \textbf{Set:} number of imputations $M$
    \State Fit parametric survival model in arm $a$, obtain maximum likelihood parameter fit $\hat{\beta}|A=a$
    \For{$\delta_j = \delta_1, \ldots, \delta_J$}
        \For{$m = 1, \ldots, M$}
            \State Sample from $\tilde{\beta} \sim MVN(\hat{\beta}, \hat{\Sigma}_{\beta})$
            \State Adjust hazard for early discontinued patients: $h'(t;\beta|t>c) = \delta_j \hat{h}(t;\tilde{\beta})$
            \State Impute survival times based on adjusted hazards
        \EndFor
    \EndFor
    \State \textbf{Output:} $M$ imputed datasets for each $\delta_j$
\end{algorithmic}
\end{algorithm}

\subsection{Ad hoc imputation}

Two alternative approaches that do not use any model for imputation are used in practice for their computational simplicity \citep{FDA2023_codebreak200}. In these ad hoc approaches, event/censoring times are imputed in one of two ways. The first involves randomly choosing a proportion of censored times to be imputed as an event \textit{at} the censoring time, and the second involves imputing all censored times by randomly sampling the $(100 \times \theta)\%$ earliest/latest event times after censoring; a form of hot-deck imputation \citep{madow1983incomplete}. See \hyperref[sec:illustration]{Appendix} Figure~\ref{fig:illustration} for a graphical illustration of these ad hoc approaches. Let $T_i$ and $C_i$ be the potential event time and right-censoring time of each $i$th patient, then $Y_i$ is the event indicator if $T_i \leq C_i$ and the observed outcome is $\min\{T_i, C_i\}$.

\subsubsection*{Landmark sampling}

Of the $N_c$ number of patients to be imputed, a random subset of $\kappa$ patients (with $0\leq\kappa\leq N_c$) will be deterministically assigned an event time at the time of censoring (i.e., $Y_i := 1$ and $C_i := T_i$) or extend the censoring time to the potential maximum follow-up of each patient (i.e., $C_i := C_i + (T_\text{data cut-off} - C_i)$), depending on the direction of imputation to have a better (for control arm) or worse (for treatment arm) extreme outcome.

At the patient level, this approach always imputes an extreme case scenario to consider patients immediately having an event (worst case) or being event free up to the longest possible follow-up time (best case), they become equivalent to the model-based imputation with individual adjustment as $\delta_i = \infty$ or $\delta_i = 0$ for patient $i$. At the trial level, when all the patients are imputed to have events at the time of censoring, the ad hoc imputation becomes equivalent to model-based imputation with $\delta \rightarrow \infty$ regardless of model choice. Conversely, when the event times of all patients are censored at the time of their respective maximum follow-up, it becomes equivalent to model-based imputation where $\delta = 0$. See Algorithm \ref{algo:model_free_deter} for a step-by-step implementation of the ad hoc tipping point with landmark sampling imputation. 

\begin{algorithm}
\caption{Landmark sampling ad hoc imputation}
\label{algo:model_free_deter}
\begin{algorithmic}[1]
    \State \textbf{Input:} survival data $(T, C, A)$ where $T$ and $C$ are event and censoring time, and $A$ is the arm assignment.
    \State \textbf{Identify:} early censored patients requiring imputation, either in the experimental or the control arm
    \State \textbf{Set:} appropriate range of $\kappa$ as the number of patients to be considered event-free at data cut-off (control arm) or having an event at censoring time (experimental arm) ($0 \leq \kappa_1, ..., \kappa_J \leq N_c$), the remaining patients are not imputed.
    \State \textbf{Set:} number of imputations $M$

    \For{$\kappa_j = \kappa_1$ to $\kappa_J$}
        \For{$m = 1$ to $M$}
            \State 
                \textbf{do} Impute $\kappa_j$ out of $N_c$ patients as having event at censoring time in experimental arm ; or
            \State 
                \textbf{do} Impute $\kappa_j$ out of $N_c$ patients as being censored at data cut-off in control arm.
        \EndFor
    \EndFor

    \State \textbf{Output:} $M$ number of imputed dataset for each $\kappa_j$
\end{algorithmic}
\end{algorithm}

\subsubsection*{Percentile sampling} 
Another ad hoc approach for imputation is to randomly sample the observed outcome from a pool of patients. The pool of patients is identified by ranking the observed times of patients who do not require imputation. That is, sample from patients with $\{i \in N | \min\{T_i, C_i\} \leq F_{\min\{T_i, C_i\}}^{-1}(\theta)\}$ for the worst $\theta$-percentile and sample from patients with $\{i \in N | \min\{T_i, C_i\} \geq F_{\min\{T_i, C_i\}}^{-1}(\theta)\}$ for the best $\theta$-percentile, where $F(.)$ is the cumulative distribution function of observed times.

Although this method does not involve model fitting, it shares two key similarities with model-based imputation. First, it assumes that patients whose event times require imputation exhibit similar behavior to the $\theta$-percentile of best or worst patients from whom they are sampled, i.e., modifying the hazard. Secondly, it is possible to accommodate categorical covariates into the sampling process by selecting the $\theta$-percentile of best/worst patients conditional on covariate values. Recent FDA ODAC and summary review utilized this percentile sampling imputation approach as sensitivity analysis \citep{fda_sotarasib, natalee}. See Algorithm \ref{algo:model_free_random} for a step-by-step implementation of the ad hoc tipping point with imputation based on percentile sampling.

\begin{algorithm}
\caption{Percentile sampling ad hoc imputation}
\label{algo:model_free_random}
\begin{algorithmic}[1]
    \State \textbf{Input:} survival data $(T, C, A)$ where $T$ and $C$ are event and censoring time, and $A$ is the arm assignment.
    \State \textbf{Identify:} early censored patients requiring imputation, either in the experimental or the control arm
    \State \textbf{Set:} appropriate range of $\theta$-percentile of best/worst event times from all observed event times ($0 < \theta_1, ...\theta_J < 100)$ 
    \State \textbf{Set:} number of imputations $M$
    \For{$\theta_j = \theta_1$ to $\theta_J$}
        \For{$m = 1$ to $M$}
            \State 
                \textbf{do} Impute by sampling from $\theta_j$-percentile of worst event times (shortest event / censoring times across both arms) in experimental arm; or
            \State
                \textbf{do} Impute by sampling from $\theta_j$-percentile of best event times (longest event / censoring times across both arms) in control arm.
        \EndFor
    \EndFor

    \State \textbf{Output:} M number of imputed dataset for each $\theta_j$
\end{algorithmic}
\end{algorithm}

\subsection{Tipping-point analysis}
Prior to tipping point analysis, a critical preliminary step is to compare key baseline characteristics between patients who required and did not require imputation, and if these characteristics differ by arm. Potential imbalances may indicate that the censoring is informative, hence highlight the potential vulnerability of the trial and underscore the importance of tipping point analyses. Of course, this can only be done for observed baseline covariates.

The primary objective of conducting tipping-point analysis is to assess which value/s of the sensitivity parameter ($\delta$ $\kappa$ or $\theta$, depending on the imputation approach) would lead to the estimated treatment effect no longer favoring the experimental arm, e.g., loss of nominal significance, or a confidence limit or point estimate crossing the null. For each sensitivity parameter value within a predefined range, multiple imputed datasets are generated and the results based on log(HRs) are combined using Rubin's rule \citep{Rubin1987}. It is recommended to specify the rules of deciding sensitivity parameter ranges a~priori, since the results of the analysis might unknowingly influence the subsequent interpretation of the sensitivity analysis \citep{Cro2020}. The range of parameter values and their corresponding results help assess the extent to which missing data could potentially affect the estimated treatment effect and whether the results remain robust under various assumptions on the missing data. The tipping point analysis identifies the parameter value where the results “tip”. This paper focuses on imputation within one arm only, but the methods are readily applied in both arms simultaneously, and sensitivity parameters can differ by arm. It is also in principle possible to use covariate-dependent sensitivity parameters to explore a broader range of scenarios.

\begin{algorithm}
\caption{Rubin's rules}
\label{algo:rubin}
\begin{algorithmic}
\For{each sensitivity parameter ($\theta_j$, $\lambda_j$, or $\kappa_j$)}
    \State \textbf{Input:} $j$th imputed dataset $(T,C,A,m)$ where $m=1,\ldots,M$ indexes imputation number
    \For{$m = 1$ to $M$}
        \State estimate $\widehat{HR}_{m,j}$
        \State estimate within-imputation variance $\widehat W_{m,j} = \widehat{\text{Var}}(HR)_{m,j}$
    \EndFor
    \State estimate pooled HR: $\widehat{HR}_j = \text{exp} \left(\frac{1}{M} \sum_{m=1}^{M} \text{log} (\widehat{HR}_{m,j}) \right)$
    \State estimate pooled variance: $\widehat{\sigma}_j^2 = \frac{1}{M} \sum_{m=1}^{M} \widehat W_m + \frac{1 + \frac{1}{M}}{M-1} \sum_{m=1}^{M} (\text{log} (\widehat{HR}_{m,j}) - \log(\widehat{HR_j}))^2$
    \State compute 95\% confidence interval (CI): $\widehat{HR}_j \times \exp\left(\pm t_{\hat{\nu},\alpha/2} \sqrt{\bar{\sigma}_j^2}\right)$, where $\hat{\nu}=(M-1)\left(1+\frac{\widehat{W}/\widehat{B}}{1+\frac{1}{M}}\right)^2$
    \EndFor
\State \textbf{Output:} dataset with $\widehat{HR}_j$, CIs and corresponding sensitivity parameter to identify tipping point (e.g. upper confidence limit crosses 1)
\end{algorithmic}
\end{algorithm}

\begin{figure}
    \centering
    \includegraphics[width=\linewidth]{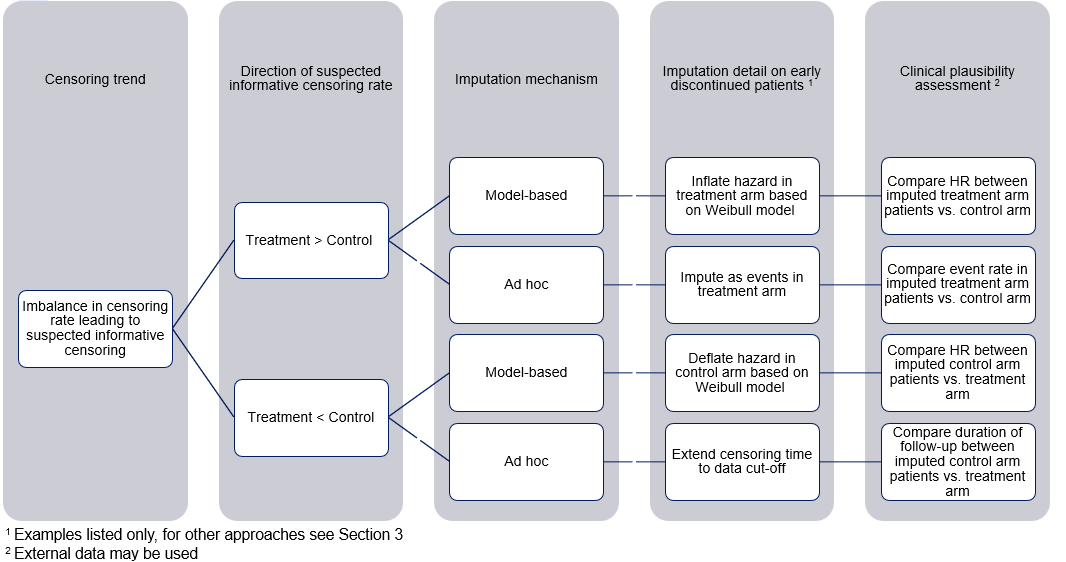}
    \caption{Flow diagram with five steps for tipping point analysis.}
    \label{fig:flow}
\end{figure}

One of the key considerations in setting up a tipping point analysis is the criteria for selecting the patients/arms to be imputed. The focus of this paper is on the imputation of the arm associated with higher censoring rate such that it may lead to an overestimation of the treatment benefit; however, all methods readily extend to imputation in multiple arms where each has a different sensitivity parameter. The choice of sensitivity parameter may vary depending on the clinical question of interest and the context of the censoring mechanism, as well as the level of conservativeness. In most cases, these choices are made such that the tipping point analysis adequately addresses the impact of the imbalance of such censoring on the robustness of the estimated treatment effect. We propose a flow diagram in Figure \ref{fig:flow} designed to guide the selection of the appropriate tipping point analysis based on different scenarios. This diagram proposes a thought process when considering statistical approaches to assess the impact of imbalance of censoring between the arms. By following this structured approach, researchers can choose the most suitable tipping point analysis method for their specific context.

Interest from regulatory agencies is often in an imputation procedure that results in the treatment effect becoming more pessimistic, as a stress-test of the original results. Consider the scenario where there is a higher study discontinuation rate in the control arm, such as in open-label studies where control arm patients with better prognosis withdraw from the trial. For ad hoc imputation, a random subset of patients who withdrew may be selected and considered event-free at the time of data cutoff, effectively extending their censoring time by $t - c$, where $t$ is the data cutoff time and $c$ is the censoring time. For model-based imputation, patients in the control arm who dropped out can be assigned a sensitivity parameter $\delta < 1$ indicating deflated hazard compared to the model fitted on patients who did not drop out. In contrast, if there is a higher study discontinuation rate in the experimental arm, a subset of patients may be imputed as having experienced an event at the time of discontinuation for the ad hoc imputation or assigned a sensitivity parameter $\delta > 1$ indicating inflated hazard for the model-based imputation. 

\subsubsection*{Links to estimand}
Let us consider the target of estimation based on treatment policy estimand, in which the time to event of interest is evaluated regardless of treatment discontinuation, initiation of subsequent therapy, or other post-randomization intercurrent events. In the presence of censoring due to study discontinuation, the event times are not known and the treatment policy estimand is then not estimable without further assumptions, such as independent censoring.

The tipping point analyses modify the assumptions used to handle missing event times following study discontinuation through imputation. Importantly, varying the sensitivity parameter or changing the post-discontinuation assumptions does not inherently alter the estimand itself. Rather, these analyses assess the robustness of inference for the same target estimand under alternative assumptions about the unobserved data mechanism.

Different imputation strategies may nevertheless resemble assumptions associated with other estimand strategies. For example, imputing an event at the time of discontinuation may conceptually resemble a composite strategy, while assuming patients remain event-free until data cut-off may resemble a hypothetical best-case scenario. However, unless such rules are systematically applied to all patients as part of the endpoint definition, these approaches do not formally redefine the estimand. Instead, they represent sensitivity analyses imposing alternative assumptions on the outcomes of patients with missing post-discontinuation data.

Some ad hoc approaches impose deterministic post-discontinuation outcomes and may therefore be difficult to justify clinically. In this context, they are best interpreted as extreme stress tests intended to evaluate the robustness of the estimated treatment effect to departures from the primary censoring-at-random assumption.

\subsubsection*{Assessing clinical plausibility of tipping points}
In addition to tipping point analysis itself, other parameter values may be useful to report, serving as aids to help assess the clinical plausibility of the tipping point results at specific values of the sensitivity parameter. Jump-to-reference imputation (\cite{Atkinson2019}) is an analysis method in which patients requiring imputation in the experimental arm are assumed to switch to the outcomes observed in the control arm (the reference) after study discontinuation. This method models the scenario where, after discontinuing the experimental treatment, patients’ outcomes behave as though they had received the standard of care from that point onward. In the delta-method for model-based imputation, $\delta$ can be selected based on an estimated hazard ratio (HR) comparing the experimental and control arms, such that the imputation reflects a scenario consistent with the jump-to-reference method. In this case, $\delta$ is set to adjust the imputed outcomes in the experimental arm to align with those expected in the control arm, simulating a switch to the standard of care after study discontinuation. In the case of ad hoc imputation, $\kappa$ or $\theta$ can be chosen such that the event rate among the $N_c$ patients who required imputations matches the event rate in the experimental or control arm. In either approach, such anchoring values (hazard ratio or event rate) help to judge how extreme an imputation procedure is. An anchoring value could alternatively be obtained from relevant published trial results, although may require additional justifications. 

To understand the overall impact of such imputation, one useful approach is to compare the Kaplan–Meier (KM) survival curves before and after imputation. A significant upward shift in the survival curve post-imputation may indicate an overly optimistic and potentially unrealistic assumption about the outcomes of control arm discontinued patients. Visualizing KM curves pre- and post-imputation is another way to evaluate the extent and plausibility of the shift caused by imputation. We further illustrate this approach in the two case studies presented in the Section \ref{sec:case}.

\section{Case studies}
\label{sec:case}
We apply model-based and ad hoc tipping point analyses to two re-constructed datasets from the studies CodeBreaK200 \citep{de2023sotorasib} and ExteNET \citep{martin2017neratinib}, to illustrate how model-based and ad hoc analyses differ in real scenarios in terms of their assumptions and their interpretations. A comparison of summary statistics between the reported and re-constructed data is shown in \hyperref[tab:compare]{Appendix} Table \ref{tab:report}, indicating fair representation of the original data. 

For the model-based tipping point analysis, We adopted a Weibull distribution for the event times, as it provides a better fit compared to an exponential distribution which assumes a constant hazard over time. The treatment effect us summarized through hazard ratios estimated using Cox proportional hazard models. The range of sensitivity parameters for each tipping point analyses are chosen such that the tipping point can be identified. Note that it is possible to not reach a tipping point at all for studies with strong treatment effect and low number of patients requiring imputations.

\subsection{CodebreaK200 analysis}
For the CodebreaK200 trial, the US FDA conducted a model-based tipping point sensitivity analysis examining the impact of asymmetric censoring and revealed that the primary PFS results were not robust when accounting for potential bias caused by early discontinued patients, with 55\% risk reduction among the early discontinued patients in the control tipping the positive results. Interestingly, the sponsor presented a different ad hoc tipping point analysis to showcase the robustness of their results while labeling the assumptions of the sensitivity analyses of the FDA as overly optimistic for the control arm. These findings were discussed during the Oncologic Drugs Advisory Committee meeting held on October 5, 2023 \citep{FDA2023_codebreak200}. In this section, we perform our own model-based and ad hoc tipping point analyses and discuss the results in relation to the ODAC findings. 

We extracted data from the published Kaplan-Meier curve in \cite{de2023sotorasib} using a digitization software. Since the timing and reasons for individual censoring were unavailable, we assumed that all censored observations within the first five weeks corresponded to patients who were censored for non-administrative reasons and likely not being followed further. This approach resulted in approximately 21 (12.1 $\%$) patients in the control arm and 6 (3.5 $\%$) patients in the experimental arm, aligning closely with the number of patients for whom the FDA imputed event times. 

\begin{figure}
    \centering
 \begin{subfigure}[t]{0.475\textwidth}
        \centering
        \includegraphics[width=\textwidth]{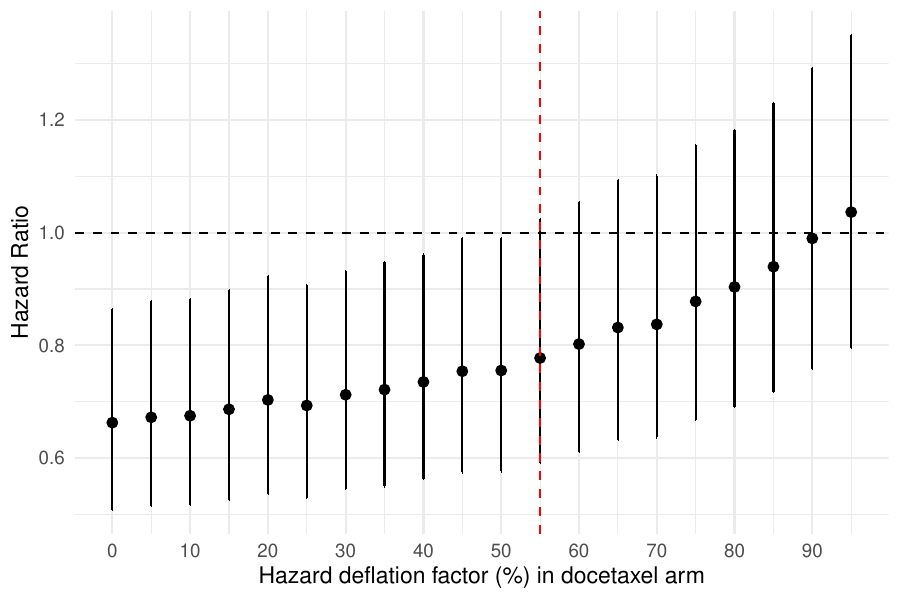} 
        \caption{Model-based tipping point analysis with range of hazard deflation ($\delta_j$) used to impute event times to early discontinued patients in control arm.}
        \label{fig:subfig3}
    \end{subfigure}
    \hfill
    \begin{subfigure}[t]{0.475\textwidth}
        \centering
        \includegraphics[width=\textwidth]{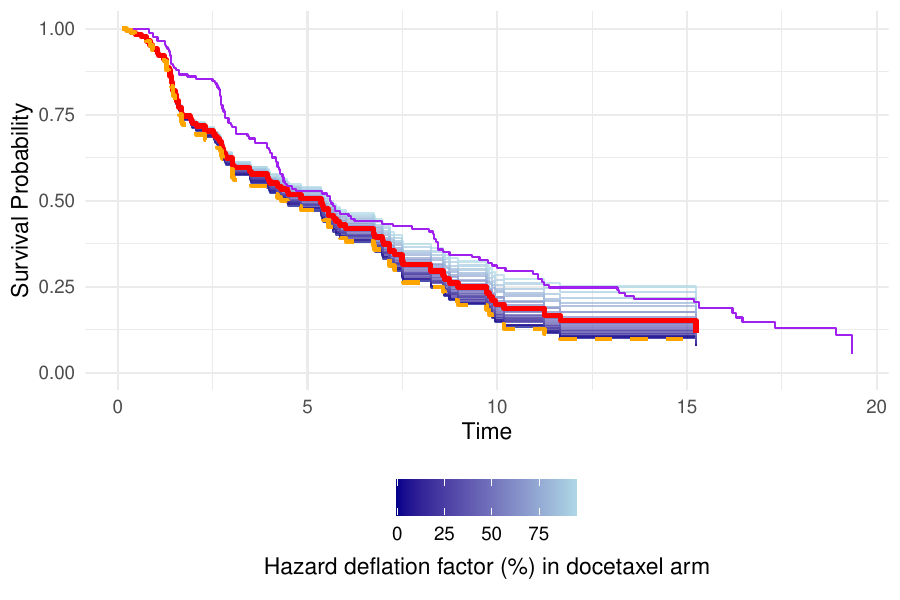} 
        \caption{See caption in Figure \ref{fig:subfig2}}
        \label{fig:subfig4}
    \end{subfigure}
    \begin{subfigure}[t]{0.475\textwidth}
        \centering
        \includegraphics[width=\textwidth]{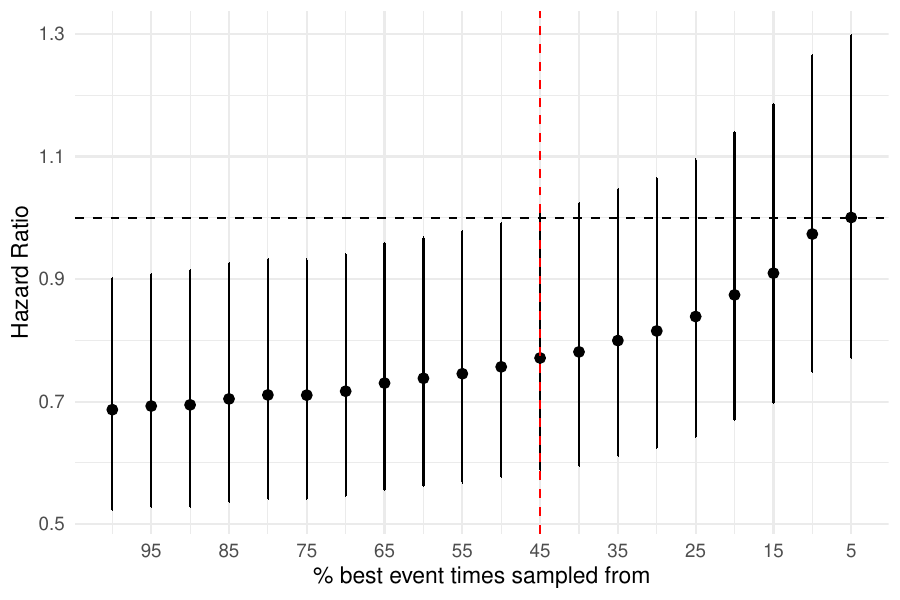} 
        \caption{Tipping point analysis using percentile sampling with different \% best event times sampled ($\theta_j$) from both arms combined for control arm imputation.}
        \label{fig:subfig1_random}
    \end{subfigure}
    \hfill
    \begin{subfigure}[t]{0.475\textwidth}
        \centering
        \includegraphics[width=\textwidth]{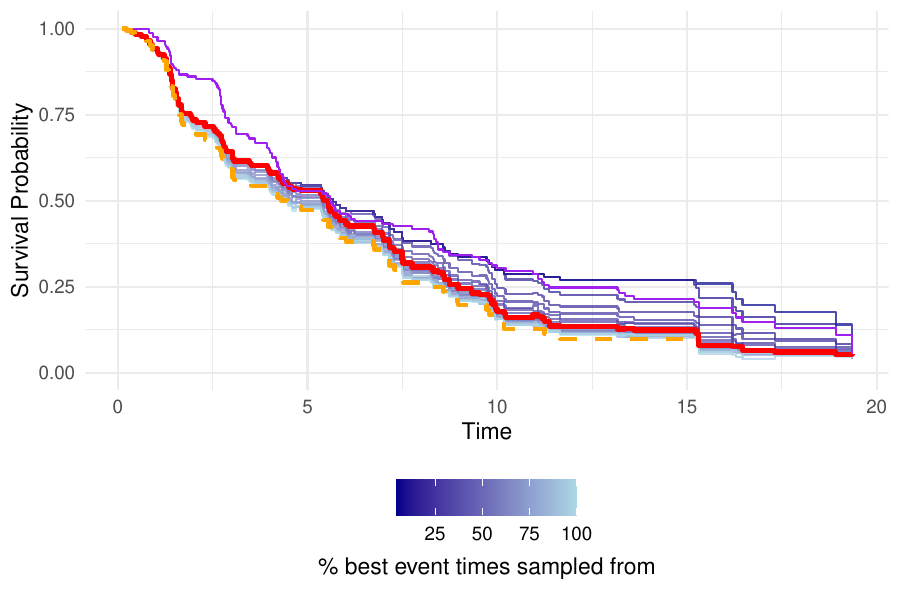} 
        \caption{See caption in Figure \ref{fig:subfig2}}
        \label{fig:subfig2_random}
    \end{subfigure}
       \begin{subfigure}[t]{0.475\textwidth}
        \centering
        \includegraphics[width=\textwidth]{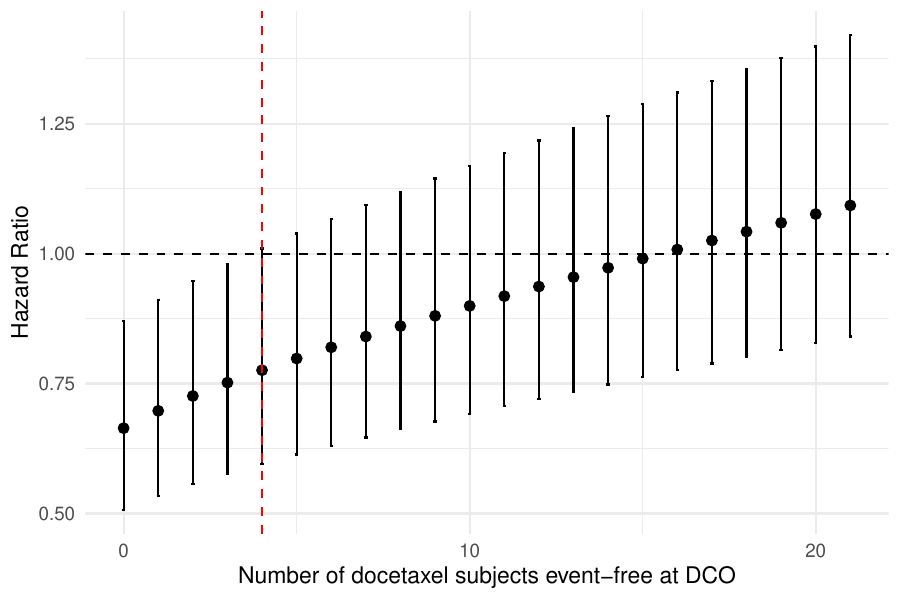} 
        \caption{Tipping point analysis using landmark sampling with different number of early discontinued patients ($\kappa_j$) in the control moved to be censored at data-cut-off.}
        \label{fig:subfig1}
    \end{subfigure}
    \hfill
    \begin{subfigure}[t]{0.475\textwidth}
        \centering
        \includegraphics[width=\textwidth]{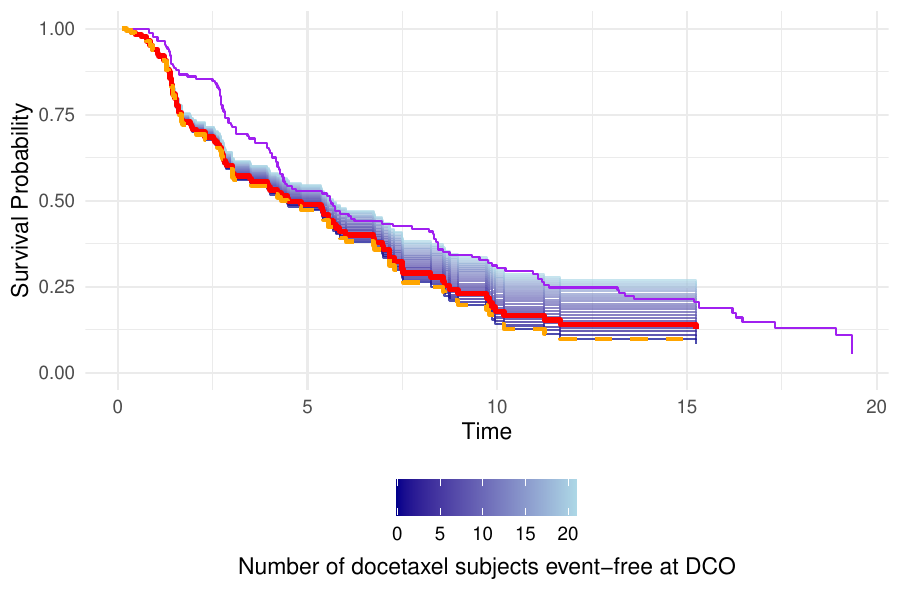} 
        \caption{KM curves for original as well as every imputed dataset are shown. The curves for the original data and the imputed data that tipped the result are shown in orange and red, respectively. The KM curve for the experimental arm (no imputation) is shown in purple.}
        \label{fig:subfig2}
    \end{subfigure}
        \caption{Tipping point analyses on re-constructed Codebreak200 data. Model-based and ad hoc imputation were applied to early discontinued patients in docetaxel arm.}
    \label{fig:codebreak200}
\end{figure}

Our results are shown in Figure \ref{fig:codebreak200}. The ad hoc approaches suggest that the results tip if at least 5 early discontinued patients in the control arm are considered as event-free at the data cut-off  or if we impute all of the early study discontinuation in the arm from a sample of 45 \% best event times (both arms combined) while the model-based approach tips the results with a hazard deflation of at least 55 $\%$ for imputing event times for early discontinued patients in the control arm. Although tipping points are expressed at the same scale as a percentage, they should not be compared against each other due to their completely different imputation mechanism. To assess the clinical plausibility on the tipping point using the trial data, 55\% hazard deflation translates to a HR of 0.45 between early discontinued patients in control arm and control arm patients who did not drop out. This translated to HR = (1 - 0.55) / 0.66 = 0.68 between the early discontinued patients in the control arm and sotorasib arm, which seems unlikely given the limited treatment options these patients have. We generate additionally the KM curves for each value of the sensitivity parameters in panel (b), (d) and (f) in Figure \ref{fig:codebreak200}. KM curves from multiply imputed datasets were pooled using Rubin’s rules after complementary log-log transformation as described in \cite{marshall2009combining}. We are particularly interested in the scenario that tips the result and the shift it causes to the original KM curve. While there is no objective measure to assess the robustness of the result, the average KM curves give a visual representation of how optimistic the assumptions of the tipping point approaches are on the control arm. In this specific case, the shift in survival curve does not seem to be overly optimistic and may point to a lack of robustness in line with FDA's findings. 

In the context of CodeBreaK200, these results should be interpreted in light of the fact that the primary PFS analysis was intended to reflect a treatment policy strategy, under which early treatment discontinuation was not considered as an intercurrent event requiring special handling. The tipping point analyses explored here therefore represent the assumption that post-discontinuation hazard on the subset of control-arm patients who were no longer followed would have been worse than those who were followed.

While all three approaches evaluate sensitivity to such departures, they do so in materially different ways: the landmark sampling approach enforces extreme assumptions at the individual level, the percentile-sampling ad hoc approach assumes similarity with favorable observed subpopulations, and the model-based approach imposes a structured, parametric modification of the post-discontinuation hazard. The tipping point stemming from the different sensitivity analyses correspond to distinct assumptions about post-discontinuation behavior and should be interpreted as probing different aspects of robustness in this trial.

\subsection{ExteNET analysis}

\begin{figure}[ht]
    \centering
        \begin{subfigure}[t]{0.475\textwidth}
        \centering
        \includegraphics[width=\textwidth]{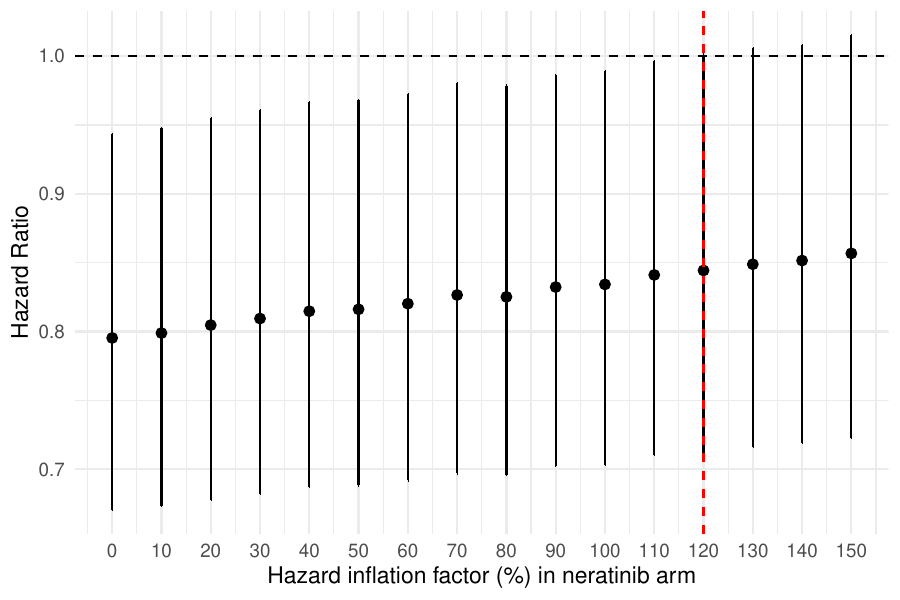} 
        \caption{Model-based tipping point analysis with range of hazard inflation ($\delta_j$) used to impute event times to early discontinued patients in neratinib arm.}
        \label{figsub:extenet_weibull}
    \end{subfigure}
    \hfill
    \begin{subfigure}[t]{0.475\textwidth}
        \centering
        \includegraphics[width=\textwidth]{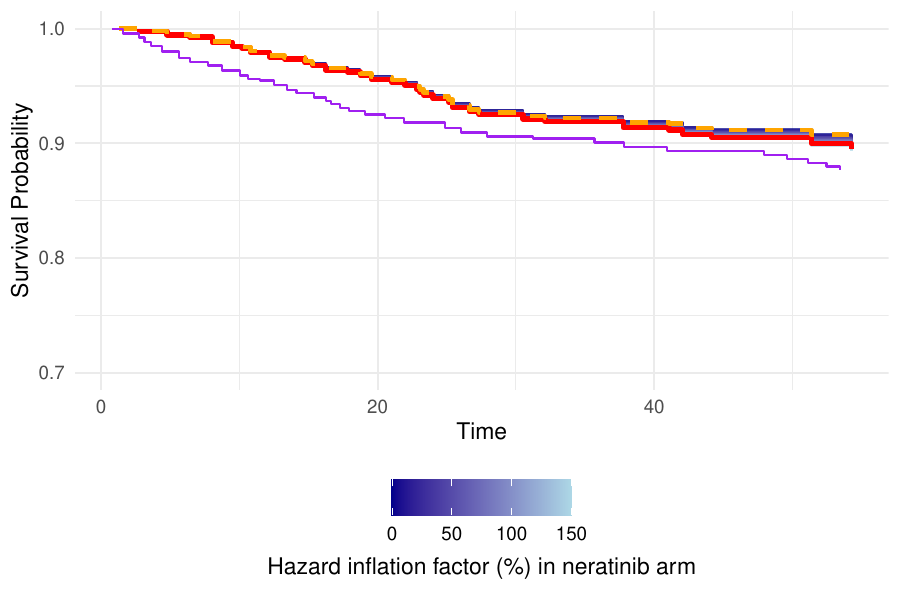} 
        \caption{KM curves for original as well as every imputed dataset are shown. The curves for the original data and the imputed data that tipped the result are shown in orange and red, respectively. The KM curve for the control arm (no imputation) is shown in purple.}
        \label{figsub:extenet_weibull_km}
    \end{subfigure}
        \begin{subfigure}[t]{0.475\textwidth}
        \centering
        \includegraphics[width=\textwidth]{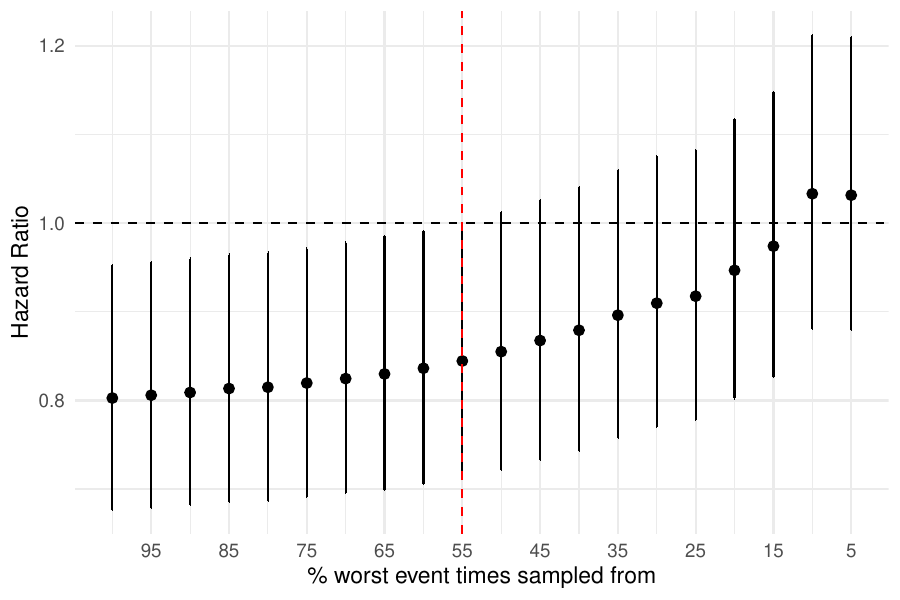} 
        \caption{Tipping point analysis using percentile sampling with different percentile of worst event times ($\theta_j$) sampled from both arms for neratinib arm imputation.}
        \label{figsub:extenet_random}
    \end{subfigure}
    \hfill
    \begin{subfigure}[t]{0.475\textwidth}
        \centering
        \includegraphics[width=\textwidth]{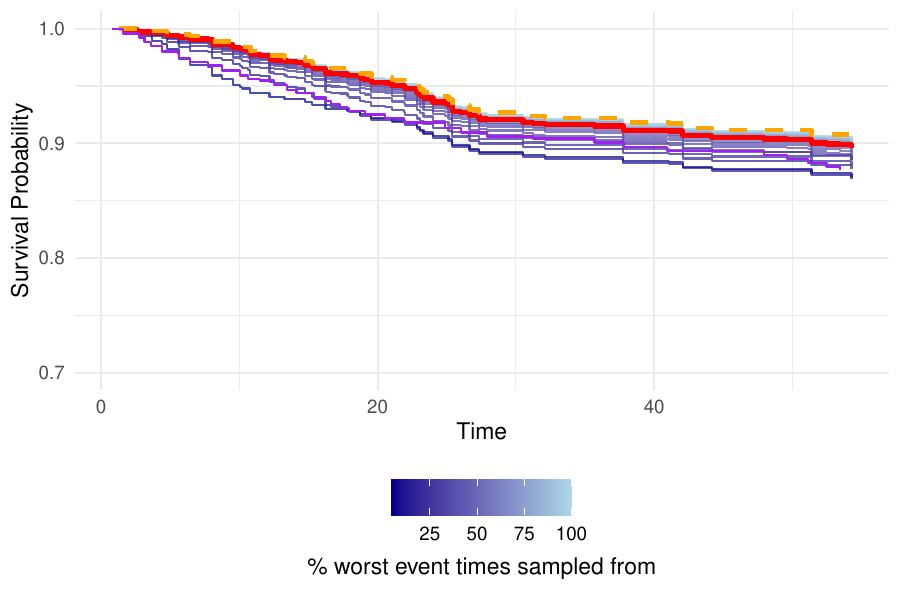} 
        \caption{See caption in Figure \ref{figsub:extenet_weibull_km}}
        \label{figsub:extenet_random_km}
    \end{subfigure}
    \caption{Tipping point analyses on re-constructed ExteNET data. Model-based and ad hoc imputation were applied to early discontinued patients in neratinib arm.}
    \begin{subfigure}[t]{0.475\textwidth}
        \centering
        \includegraphics[width=\textwidth]{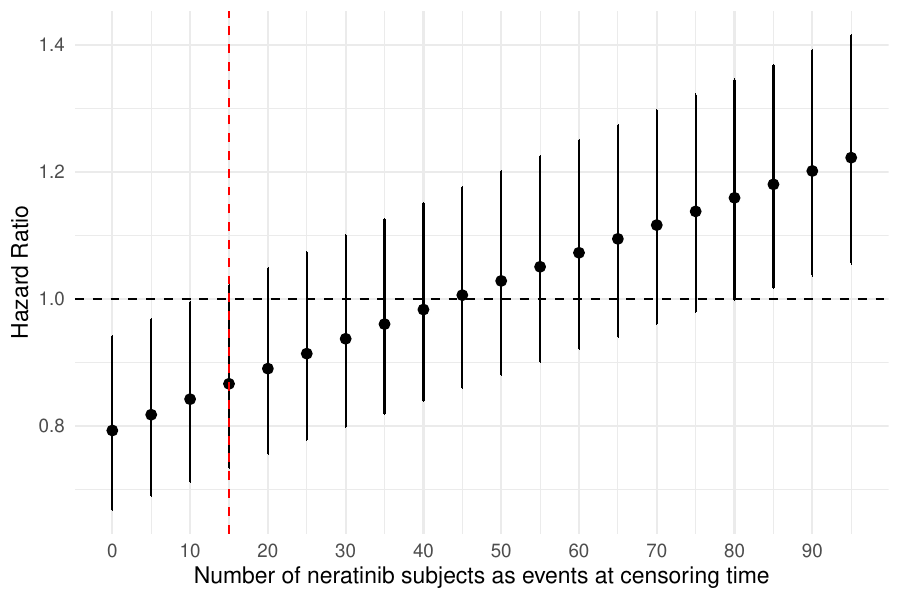} 
        \caption{Tipping point analysis using landmark sampling with different number of early discontinued patients ($\kappa_j$) in the neratinib arm considered as having an event at time of study discontinuation.}
        \label{figsub:extenet_deterministic}
    \end{subfigure}
    \hfill
    \begin{subfigure}[t]{0.475\textwidth}
        \centering
        \includegraphics[width=\textwidth]{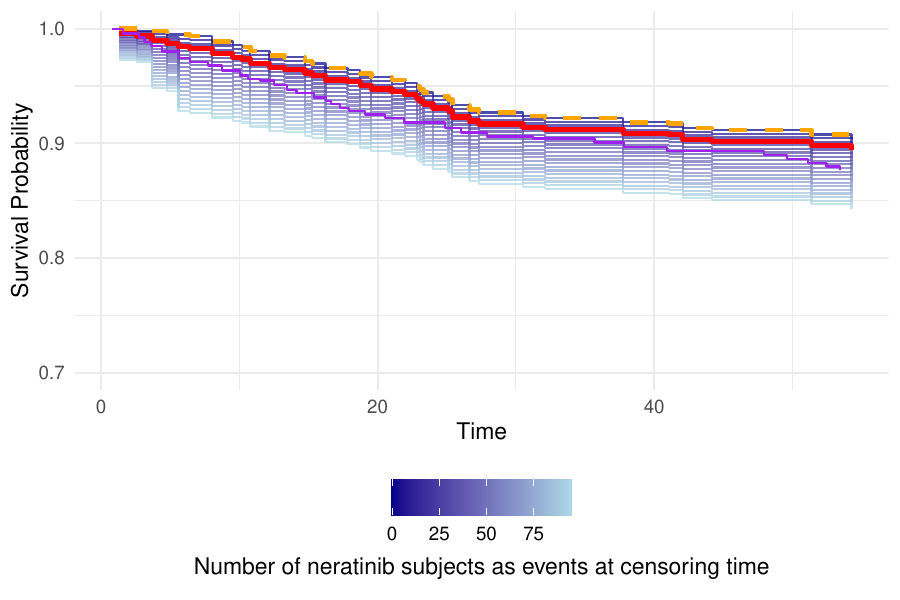} 
        \caption{See caption in Figure \ref{figsub:extenet_weibull_km}}
        \label{figsub:extenet_deterministic_km}
    \end{subfigure}
        \label{fig:ExteNET}
\end{figure}

Similar to the previous case study, we digitized the published Kaplan–Meier (KM) curve from the ExteNET trial \citep{martin2017neratinib} using a KM digitizer and applied tipping point analyses to the re-constructed patient-level data. Patients censored within approximately the first six months were considered as early discontinuations and selected for imputation. Because early study discontinuation was more frequent in the neratinib arm, imputations were applied to this arm in a direction that worsens outcomes. The results of the different tipping point approaches are summarized in Figure \ref{fig:ExteNET}.

Under the model-based approach, imputing event times for early discontinued patients in the experimental arm using a Weibull model required a hazard inflation of 120\% to reach the tipping point. The changes in HRs are less sensitive to the tipping point parameter (Figure \ref{figsub:extenet_weibull}) compared to the CodebreaK200 case study, as a result of a smaller percentage of discontinued patients and therefore a smaller impact from the imputations on the overall results. 

In contrast, the landmark sampling approach, which imputes events at the time of study discontinuation, represents the most conservative assumption at the individual patient level. For patients with early-stage breast cancer, assuming that study discontinuation within six months corresponds immediately to progression constitutes an extremely pessimistic scenario that may not align with clinical reality. This approach results in large shifts in the KM curves and a broader range of HR estimates compared to the model-based method using hazard inflation. The results were tipped when at least 15 discontinued patients in the neratinib arm are assumed to have events imputed at the time of study discontinuation. As for the percentile sampling approach, where we sample event times from the observed data, sampling from the 55th-percentile worst event times (both arms combined) tips the results. 

We emphasize that direct comparisons across tipping point approaches are not meaningful, as the sensitivity parameters ($\kappa, \lambda, \theta$) operate on fundamentally different scales and reflect distinct imputation mechanisms. Nevertheless, the KM curves in Figures \ref{figsub:extenet_weibull_km}, \ref{figsub:extenet_deterministic_km}, and \ref{figsub:extenet_random_km} show that, at the tipping point, each approach produces a similar overall shift by definition since it is where the upper bound of the 95\% confidence interval for the HR crosses unity. Rather than focusing on the numerical value of the sensitivity parameter itself, interpretation should rely on clinical plausibility grounded in the trial context. For example, a 120\% hazard inflation – equivalent to more than doubling the hazard rate for discontinued patients in the experimental arm – is required to tip the results. This corresponds to an estimated hazard rate for early discontinued patients in the neratinib arm being almost three times higher than that of the control arm, with $\widehat{\text{HR}} = (1 + 1.2)/0.72 = 3.06$. Such a scenario appears highly improbable, considering the favorable benefit–risk profile of neratinib.

\section{Discussion}
Tipping point analysis is a useful statistical method for sensitivity analysis of time-to-event endpoints when informative censoring is suspected. In case of suspected informative censoring, such analyses are increasingly requested as part of regulatory review to assess the robustness of estimated treatment effect in the main analysis of clinical trials, providing additional evidence to support the benefit-risk assessment of an experimental drug. In this paper, we have described three methods to conduct these analyses and interpret their results.

Before conducting any tipping point analysis, an essential first step is to compare key baseline characteristics between patients who discontinued the study and those who did not in each arm. If meaningful differences are observed – such as in disease severity, comorbidities, or other prognostic factors – this suggests that patients who discontinued may have systematically different outcomes than those who remained, rendering the trial susceptible to informative censoring. In such settings, these observed discrepancies can help guide both the choice of imputation mechanism and the clinical plausibility of pessimistic or optimistic assumptions explored within a tipping point framework. In particular, meaningful imbalances between arms in prognostic covariates may strongly influence the clinical plausibility of the hazard functions used to impute missing event times, as the post-discontinuation risk profile of these patients may differ from that of patients remaining under observation. Accounting for such differences is therefore an important consideration when specifying sensitivity scenarios. Model-based tipping point approaches provide a natural framework to incorporate such information, as they allow the inclusion of auxiliary (endogenous or exogenous) variables in the imputation process. By conditioning the imputation model on relevant prognostic covariates, these approaches can generate post-discontinuation event trajectories that more closely reflect clinically plausible patient risk profiles. Importantly, however, the absence of observable differences does not imply independent censoring. Study discontinuation may still depend on unmeasured or unobserved aspects of prognosis, and therefore the potential for informative censoring cannot be excluded solely on the basis of similar observed characteristics.

In this paper, we classified tipping point analysis seen in practice into two broad categories based on their data imputation mechanism, namely model-based and ad hoc approaches. We recall that they rely on fundamentally different assumptions about the censoring mechanism and the underlying event time distribution. Model-based approaches typically preserve the structure of the primary analysis model and introduce sensitivity parameters within a coherent statistical framework. In contrast, ad hoc approaches often impose direct modifications to observed outcomes or summary measures without explicitly modeling the event time process. While such approaches may be simple to implement and transparent, the assumptions they encode can be difficult to justify clinically and may, in some settings, be regarded as implausible representations of the underlying data-generating mechanism. Conventionally, the tipping point results are often used to evaluate the robustness of the observed efficacy, i.e., an extreme sensitivity parameter required to tip the result indicates high probability of strong HR favoring the experimental arm. Additional factors that impact the tipping point results also include imputation approach, model specification, number of patients requiring imputation and the relative timing of censoring. As a result, the magnitude of a tipping point should not be interpreted in isolation, but rather in the context of the assumptions it encodes about post-discontinuation outcomes. Taking these factors into account can improve the reliability and interpretability of sensitivity analyses and support more informed regulatory and clinical decision-making. 

Using the case studies based on CodeBreak200 and ExteNET, we conducted an in-depth illustration of the differences between model-based and ad hoc tipping point analyses. For the CodeBreak200 case study, the ad hoc and model-based approaches captured a comparable range of resulting hazard ratios. Since their sensitivity parameters are on a different scale, direct comparisons are not feasible. The two tipping point approaches rather provide complementary assessments of the robustness of efficacy under different assumptions. They align only in the most extreme scenarios. Specifically, they become equivalent when there is a higher study discontinuation rate in the experimental arm, with $\delta \rightarrow \infty$ in the model-based method, while all discontinued patients are imputed as events in the ad hoc approach. Conversely, equivalence occurs when the imbalance favors the control arm, $\delta \rightarrow 0$, and all discontinued patients in the control arm are imputed as event-free at the maximum follow-up. The clinical plausibility assessment step provides clinical interpretation of the tipping point results, which can be used to assess the likelihood of different scenarios against external data. The reporting of multiple-imputed Kaplan-Meier curves from the tipping point analyses illustrate the magnitude of change introduced when conducting imputation, offering an alternative presentation of the tipping point results. In the model-based method, the hazard is either inflated or deflated to impute event times, which in turn affects the shape of the Kaplan-Meier curves by modifying the survival probability estimates in both arms. On the other hand, the ad hoc approach, by imputing discontinued patients as either events or event-free at maximum follow-up time, will also shift the KM curves, but the interpretation of these shifts is driven more by the assumption of how discontinued patients behave rather than adjusting for their exact timing. Although the approaches are not directly comparable numerically, comparing the resulting KM curve shifts provides a common visual framework to understand how each method perturbs the observed data.

In practice, there is rarely a single preferred tipping point method. Employing multiple approaches can provide a more comprehensive assessment of robustness by examining sensitivity across a broader range of assumptions. The imputation approaches are closely linked with the estimand thinking \citep{ICH_E9_R1_2019, Siegel2024}. The key interest in the regulatory context is to conduct sensitivity analysis for the same estimand as the main analysis \cite{Cesar2025}, which often targets treatment policy estimand. In the presence of potentially informative censoring, the assumption for standard analyses methods in time-to-event endpoints are violated to an unknown extent. Tipping point analysis explores this unknown extent to evaluate the robustness of the estimated treatment effect.

Tipping point analysis is most impactful when a clear clinical interpretation can be assigned to the scenario at which the tipping point occurs. For instance, in our Codebreak200 case study, the tipping point was reached at $\delta = 55\%$, approximately corresponding to a hazard ratio of $0.68$ between the early discontinued patients in the control arm and sotorasib arm. This translates to a meaningfully more favorable outcome for the patients who dropped out compared to the sotorasib arm making it an improbable scenario in clinical practice, as patients who dropped out were unlikely to access treatments more effective than sotorasib. Importantly, this clinical plausibility assessment step need not rely solely on trial data; comparisons with external data from the literature can also provide valuable insights.

In this paper, we focused on scenarios involving two-arm trials. However, in studies with more than two arms, the model-based approach can leverage the additional data for model fitting, potentially resulting in more precise quantification of the tipping point. This could be particularly useful in capturing nuanced differences between multiple experimental arms. 
In non-inferiority trials, the choice of imputation parameters and the specific patient groups to be imputed may differ significantly from those in superiority trials. Non-inferiority settings often demand more stringent clinical plausibility assessment of the tipping point, as the clinical implications hinge on demonstrating that the treatment is not unacceptably worse than the comparator, rather than outright superiority. In addition, while the examples presented here consider varying the sensitivity parameter in a single arm such that the treatment effect is no longer in favor of the investigational arm, the framework extends naturally to a two-dimensional setting in which sensitivity parameters are varied simultaneously across both arms. Such an extension may be relevant in clinical contexts where the reasons for study discontinuation differ between treatment groups; for example, patients discontinuing in the control arm may have a more favorable prognosis or better access to subsequent therapies, whereas patients discontinuing in the experimental arm may do so due to treatment-related toxicity. In this setting, it may be clinically reasonable to impose optimistic post-discontinuation assumptions in the control arm while simultaneously adopting more pessimistic assumptions in the experimental arm. Visualization of such scenarios can be achieved using heatmaps over combinations of sensitivity parameters, potentially yielding multiple tipping points. The general considerations regarding interpretation, clinical plausibility, and alignment with the treatment policy estimand discussed in this paper extend naturally to this more general setting.

Regardless of the chosen approach, the proportional hazards assumption no longer holds following imputation. For this reason, alternative population-level summary measures, such as restricted mean survival time (RMST) \citep{royston2013restricted}, can provide a robust and assumption-lean metric to evaluate when the tipping point occurs. RMST comparisons are particularly valuable as an alternative estimand in scenarios where proportional hazards are unlikely to hold, offering a more interpretable assessment of clinical impact. One other extension of the tipping point analysis described in our paper is the incorporation of baseline covariates to enable the imputation tailoring to different patient groups \citep{Jin2024cov}.  

In summary, tipping point analysis provides a practical tool to assess the robustness of efficacy results under potential informative censoring. To support reliable interpretation, we encourage transparent reporting, routine visualization, and clinical plausibility assessment, including key drivers such as the imputation model and the characteristics of censored patients requiring imputation. An open-source implementation of the various approaches described in the manuscript has been made available via an R package \textbf{tipse} \citep{tipse}. We believe consistent implementation, transparent reporting and clinical plausibility assessment of the tipping point against internal and external data together improves the overall reliability of tipping point analyses.  

\section*{Acknowledgments}
The authors would like to thank Dr Jiawei Wei (Novartis) and two anonymous reviewers from Pharmaceutical Statistics for their insightful feedback on earlier versions of the manuscript.

\section*{Conflict of interest}
Authors declare that they have no conflicts of interest.

\bibliographystyle{abbrvnat}
\bibliography{references}

@article{Jackson2014,
    author = {Jackson, Dan and White, Ian R. and Seaman, Shaun and Evans, Hannah and Baisley, Kathy and Carpenter, James},
    title = {Relaxing the independent censoring assumption in the Cox proportional hazards model using multiple imputation},
    journal = {Statistics in Medicine},
    volume = {33},
    number = {27},
    pages = {4681-4694},
    doi = {https://doi.org/10.1002/sim.6274},
    url = {https://onlinelibrary.wiley.com/doi/abs/10.1002/sim.6274},
    eprint = {https://onlinelibrary.wiley.com/doi/pdf/10.1002/sim.6274},
    year = {2014}
}

@article{schafer_graham_2002,
  author  = {Schafer, Joseph L. and Graham, John W.},
  title   = {Missing Data: Our View of the State of the Art},
  journal = {Psychological Methods},
  year    = {2002},
  volume  = {7},
  number  = {2},
  pages   = {147--177},
  doi     = {10.1037/1082-989X.7.2.147}
}

@article{sterne_2009,
  author  = {Sterne, Jonathan A. C. and White, Ian R. and Carlin, John B. and
             Spratt, Michael and Royston, Patrick and Kenward, Michael G. and
             Wood, Angela M. and Carpenter, James R.},
  title   = {Multiple imputation for missing data in epidemiological and clinical research: potential and pitfalls},
  journal = {BMJ},
  year    = {2009},
  volume  = {338},
  pages   = {b2393},
  doi     = {10.1136/bmj.b2393}
}

@Manual{tipse,
    title = {tipse: Tipping Point Analysis for Survival Endpoints},
    author = {Ajmal Oodally and Craig Wang},
    year = {2026},
    note = {R package version 1.2},
  }

@article{Meng1994,
  title = {Multiple-Imputation Inferences with Uncongenial Sources of Input},
  volume = {9},
  ISSN = {0883-4237},
  url = {http://dx.doi.org/10.1214/ss/1177010269},
  DOI = {10.1214/ss/1177010269},
  number = {4},
  journal = {Statistical Science},
  publisher = {Institute of Mathematical Statistics},
  author = {Meng, Xiao-Li},
  year = {1994},
  month = nov 
}

@book{Rubin1987,
    author = {Rubin, D.},
    title= {Multiple imputation for nonresponse in surveys},
    year = {1987},
    doi = {10.1002/9780470316696},
    address = {New York},
    publisher = {John Wiley and Sons},
}

@article{Cro2020,
    author = {Cro, Suzie and Morris, Tim P. and Kenward, Michael G. and Carpenter, James R.},
    title = {Sensitivity analysis for clinical trials with missing continuous outcome data using controlled multiple imputation: A practical guide},
    journal = {Statistics in Medicine},
    volume = {39},
    number = {21},
    pages = {2815-2842},
    doi = {https://doi.org/10.1002/sim.8569},
    url = {https://onlinelibrary.wiley.com/doi/abs/10.1002/sim.8569},
    year = {2020}
}

@incollection{madow1983incomplete,
  author={Ford, B},
  booktitle={Incomplete Data in Sample Surveys: Theory and bibliographies},
  title={An Overview of HotDeck Procedures},
  editor={Madow, W.G. and Nisselson, H. and Olkin, I. and Rubin, D.B.},
  isbn={9780123639028},
  series={Incomplete Data in Sample Surveys},
  year={1983},
  publisher={Academic Press}
}

@article{rubin1976inference,
  title={Inference and missing data},
  author={Rubin, Donald B},
  journal={Biometrika},
  volume={63},
  number={3},
  pages={581--592},
  year={1976},
  publisher={Oxford University Press}
}

@article{Cesar2025,
author = {Torres, Cesar and Levin, Gregory and Rubin, Daniel and Koh, William and Chiu, Rebecca and Permutt, Thomas},
title = {A Tipping Point Method to Evaluate Sensitivity to Potential Violations in Missing Data Assumptions},
journal = {Pharmaceutical Statistics},
volume = {24},
number = {3},
pages = {e70002},
doi = {https://doi.org/10.1002/pst.70002},
year = {2025}
}

@article{Fallah2024,
author = {Fallah, Jaleh and Mulkey, Flora and Fiero, Mallorie H. and Gittleman, Haley and Song, Chi and Puthiamadathil, Jeevan and Amatya, Anup and Agrawal, Sundeep and Vellanki, Paz and Suzman, Daniel L. and Singh, Harpreet and Amiri-Kordestani, Laleh and Mishra-Kalyani, Pallavi and Pazdur, Richard and Kluetz, Paul G. },
title = {Equipoise Lost? Trial Conduct Challenges in an Era of Breakthrough Therapies},
journal = {Journal of Clinical Oncology},
volume = {0},
number = {0},
pages = {JCO-24-01200},
year = {2024},
doi = {10.1200/JCO-24-01200},

note ={PMID: 39288354},

URL = {https://ascopubs.org/doi/abs/10.1200/JCO-24-01200},
eprint = {https://ascopubs.org/doi/pdf/10.1200/JCO-24-01200}

}

@article{Siegel2024,
author = {Siegel, Jonathan M. and Weber, Hans-Jochen and Englert, Stefan and Liu, Feng and Casey, Michelle and the Pharmaceutical Industry Working Group on Estimands in Oncology},
title = {Time-to-event estimands and loss to follow-up in oncology in light of the estimands guidance},
journal = {Pharmaceutical Statistics},
volume = {23},
number = {5},
pages = {709-727},
doi = {https://doi.org/10.1002/pst.2386},
url = {https://onlinelibrary.wiley.com/doi/abs/10.1002/pst.2386},
eprint = {https://onlinelibrary.wiley.com/doi/pdf/10.1002/pst.2386},
year = {2024}
}

@article{Lipkovich2016,
author = {Lipkovich, Ilya and Ratitch, Bohdana and O'Kelly, Michael},
title = {Sensitivity to censored-at-random assumption in the analysis of time-to-event endpoints},
journal = {Pharmaceutical Statistics},
volume = {15},
number = {3},
pages = {216-229},
doi = {https://doi.org/10.1002/pst.1738},
url = {https://onlinelibrary.wiley.com/doi/abs/10.1002/pst.1738},
eprint = {https://onlinelibrary.wiley.com/doi/pdf/10.1002/pst.1738},
year = {2016}
}

@article{Atkinson2019,
author = {Atkinson, Andrew and Kenward, Michael G. and Clayton, Tim and Carpenter, James R.},
title = {Reference-based sensitivity analysis for time-to-event data},
journal = {Pharmaceutical Statistics},
volume = {18},
number = {6},
pages = {645-658},
doi = {https://doi.org/10.1002/pst.1954},
url = {https://onlinelibrary.wiley.com/doi/abs/10.1002/pst.1954},
year = {2019}
}

@article{Jin2024cov,
title = {Imputation methods for informative censoring in survival analysis with time dependent covariates},
journal = {Contemporary Clinical Trials},
volume = {136},
pages = {107401},
year = {2024},
issn = {1551-7144},
doi = {https://doi.org/10.1016/j.cct.2023.107401},
url = {https://www.sciencedirect.com/science/article/pii/S1551714423003245},
author = {Man Jin}
}

@article{de2023sotorasib,
  title={Sotorasib versus docetaxel for previously treated non-small-cell lung cancer with KRASG12C mutation: a randomised, open-label, phase 3 trial},
  author={de Langen, Adrianus Johannes and Johnson, Melissa L and Mazieres, Julien and Dingemans, Anne-Marie C and Mountzios, Giannis and Pless, Miklos and Wolf, J{\"u}rgen and Schuler, Martin and Lena, Herv{\'e} and Skoulidis, Ferdinandos and others},
  journal={The Lancet},
  volume={401},
  number={10378},
  pages={733--746},
  year={2023},
  publisher={Elsevier}
}

@article{martin2017neratinib,
  title={Neratinib after trastuzumab-based adjuvant therapy in HER2-positive breast cancer (ExteNET): 5-year analysis of a randomised, double-blind, placebo-controlled, phase 3 trial},
  author={Martin, Miguel and Holmes, Frankie A and Ejlertsen, Bent and Delaloge, Suzette and Moy, Beverly and Iwata, Hiroji and von Minckwitz, Gunter and Chia, Stephen KL and Mansi, Janine and Barrios, Carlos H and others},
  journal={The lancet oncology},
  volume={18},
  number={12},
  pages={1688--1700},
  year={2017},
  publisher={Elsevier}
}

@article{marshall2009combining,
  title={Combining estimates of interest in prognostic 446 modelling studies after multiple imputation: current practice and guidelines},
  author={Marshall, A and Altman, D and Holder, R and Royston, P},
  journal={BMC Med Res 447 Methodol},
  volume={9},
  number={57},
  pages={448},
  year={2009}
}

@article{sui2023application,
  title={Application of Tipping Point Analysis in Clinical Trials using the Multiple Imputation Procedure in SAS},
  author={Sui, Yunxia and Bu, Xianwei and Li, Yihan and Wang, Xin},
  year={2023},
  journal={PharmaSUG Baltimore}
}

@article{gorst2022fast,
  title={Fast tipping point sensitivity analyses in clinical trials with missing continuous outcomes under multiple imputation},
  author={Gorst-Rasmussen, Anders and Tarp-Johansen, Mads Jeppe},
  journal={Journal of Biopharmaceutical Statistics},
  volume={32},
  number={6},
  pages={942--953},
  year={2022},
  publisher={Taylor \& Francis}
}

@article{yan2009missing,
  title={Missing data handling methods in medical device clinical trials},
  author={Yan, Xu and Lee, Shiowjen and Li, Ning},
  journal={Journal of Biopharmaceutical Statistics},
  volume={19},
  number={6},
  pages={1085--1098},
  year={2009},
  publisher={Taylor \& Francis}
}

@article{kleinbaum2012parametric,
  title={Parametric survival models},
  author={Kleinbaum, David G and Klein, Mitchel and Kleinbaum, David G and Klein, Mitchel},
  journal={Survival Analysis: A Self-Learning Text},
  pages={289--361},
  year={2012},
  publisher={Springer}
}

@article{liublinska2014sensitivity,
  title={Sensitivity analysis for a partially missing binary outcome in a two-arm randomized clinical trial},
  author={Liublinska, Victoria and Rubin, Donald B},
  journal={Statistics in medicine},
  volume={33},
  number={24},
  pages={4170--4185},
  year={2014},
  publisher={Wiley Online Library}
}

@misc{FDA2023_codebreak200,
  author = {AMGEN},
  title = {Oncologic Drugs Advisory Committee Meeting},
  year = {2023},
  month = {October 5},
  url = {https://www.fda.gov/media/172757/download}
}

@misc{ICH_E9_R1_2019,
  author       = {ICH},
  title        = {ICH E9 (R1) Addendum on Estimands and Sensitivity Analysis in Clinical Trials to the Guideline on Statistical Principles for Clinical Trials},
  year         = {2019},
  url          = {https://www.ich.org/page/efficacy-guidelines}
}

@article{royston2013restricted,
  title={Restricted mean survival time: an alternative to the hazard ratio for the design and analysis of randomized trials with a time-to-event outcome},
  author={Royston, Patrick and Parmar, Mahesh KB},
  journal={BMC medical research methodology},
  volume={13},
  pages={1--15},
  year={2013},
  publisher={Springer}
}

@misc{natalee,
	author = {US FDA},
	title = {{N}{D}{A}/{B}{L}{A} {M}ultidisciplinary {R}eview and {E}valuation},
	howpublished = {\url{https://www.accessdata.fda.gov/drugsatfda_docs/nda/2024/209092s018,209935s027MultidisciplineR.pdf}},
	year = {2024},
	note = {[Page 77 - 78. Accessed 18-03-2025]},
}

@misc{fda_sotarasib,
	author = {US FDA},
	title = {{O}ctober 5, 2023: {M}eeting of the {O}ncologic {D}rugs {A}dvisory {C}ommittee --- fda.gov},
	howpublished = {\url{https://www.fda.gov/advisory-committees/advisory-committee-calendar/october-5-2023-meeting-oncologic-drugs-advisory-committee-meeting-announcement-10052023}},
	year = {2023},
	note = {[Page 75 - 87. Accessed 18-03-2025]},
}

@article{little2012prevention,
  title={The prevention and treatment of missing data in clinical trials},
  author={Little, Roderick J and D'agostino, Ralph and Cohen, Michael L and Dickersin, Kay and Emerson, Scott S and Farrar, John T and Frangakis, Constantine and Hogan, Joseph W and Molenberghs, Geert and Murphy, Susan A and others},
  journal={New England Journal of Medicine},
  volume={367},
  number={14},
  pages={1355--1360},
  year={2012},
  publisher={Mass Medical Soc}
}

\section*{Appendix}
\label{sec:app}

\subsection*{Comparison of reported and re-constructed case study results}
Table \ref{tab:report} presents an evaluation of the re-constructed versus actual study data for CodeBreak200 and ExteNET trials. The comparison between the reported quantities showed those summaries  derived from the reconstruction demonstrates the high resemblance of the actual study results. 

\begin{table}[ht]
    \centering
    \begin{tabular}{l|c|c|c|c} \hline 
       Study&  \multicolumn{2}{|c|}{CodebreaK200}&  \multicolumn{2}{|c}{ExteNET}\\ \hline 
 Arm& Sotarasib& Control& Neratinib&Control
\\ \hline & \multicolumn{4}{|c}{\textbf{Reported}}\\
       \hline
       HR (95\%CI) &   \multicolumn{2}{|c|}{0.66 (0.51, 0.86)} &  \multicolumn{2}{|c}{0.73 (0.57, 0.92)} \\ \hline 
              Median (95\% CI)  & 5.6 (4.3-7.8)  & 4.5 (3.0-5.7) & NA & NA
\\ \hline 
 Number of events (\%)& 122 (71.3) & 101 (58.0) & 116 (8.2) & 163 (11.5) 
\\ \hline 
       Number of study discontinuation (\%)  &   2 (1.2)& 23 (13.2)& 80 (5.6) & 25 (1.8) 
\\ \hline & \multicolumn{4}{|c}{\textbf{Re-constructed}}\\
       \hline 
       HR (95\%CI) &  \multicolumn{2}{|c|}{0.66 (0.51, 0.86)} & \multicolumn{2}{|c}{0.72 (0.56-0.92)} \\ \hline 
              Median (95\% CI)  & 5.6 (4.3-8.2) & 4.5 (3.0-5.9) & NA & NA
\\ \hline 
 Number of events (\%)& 124 (72.5) & 105 (60.3) & 115 (8.1) & 159 (11.2)
\\ \hline 
       Number of study discontinuation (\%)  & 6 (3.5) & 21 (12.1)& 97 (6.8) & 27 (1.9)
\\ \hline 
    \end{tabular}
    \caption{Summary of reported and re-constructed case study results.}
    \label{tab:report}
\end{table}

\subsection*{Simulation setup for the simulated examples in Section 2.3}
\label{sec:sim}
We simulate trials with $N = 2,000$ randomized into either treatment or control arm ($A = 0, 1$) in 1:1 ratio under simple randomization. We denote the time to event and censoring as $T_i$ and $C_i$ for the $i$th patient, respectively. The event times are simulated under Weibull distribution conditional on the arm $A$ and the prognostic covariate $X$. The hazard function for the event is expressed as the following with proportional hazards:
$$ h_i(t) = \gamma \lambda (t^{\gamma-1}) \exp(\log(0.7) A_i + \log(0.75) X_i) $$
where $X \sim N(0, 1)$ is a continuous prognostic covariate. At the same time, a censoring mechanism is introduced separately using the same prognostic covariate with treatment interaction to create the scenario of informative censoring and imbalance in study discontinuation. The hazard function for censoring is based on a conditional exponential distribution in two different settings:
\begin{itemize}
     \item[1.] For higher censoring rate in the experimental arm patients with worse prognosis:
     $$ h_i(c) = \lambda \exp(aA_i + \log(0.5) X_i + \log(0.5) A_i X_i) $$
     \item[2.] For higher censoring rate in the control arm patients with better prognosis:
     $$ h_i(c) = \lambda \exp(aA_i + \log(2) X_i + \log(2) A_i X_i)$$
 \end{itemize}
where $a$ is varied across simulation settings to result in different level of imbalance in the censoring rate. 

\newpage
\subsection*{Illustration of ad hoc imputations}
\label{sec:illustration}
\begin{figure}[ht]
    \centering
    \begin{subfigure}[t]{\textwidth}
    \includegraphics[width=\linewidth]{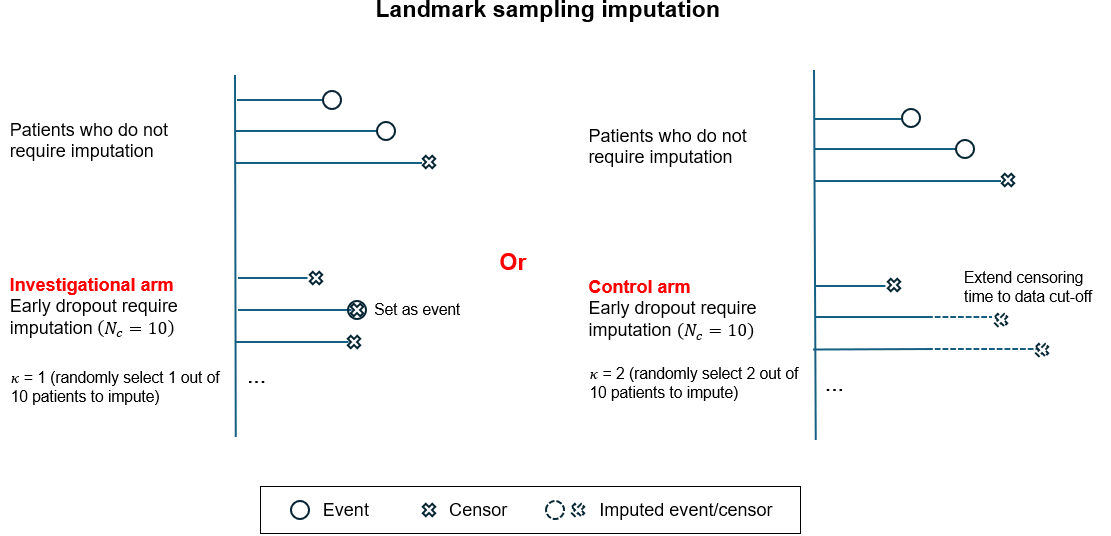}
    \end{subfigure}
    \begin{subfigure}[t]{\textwidth}
    \includegraphics[width=\linewidth]{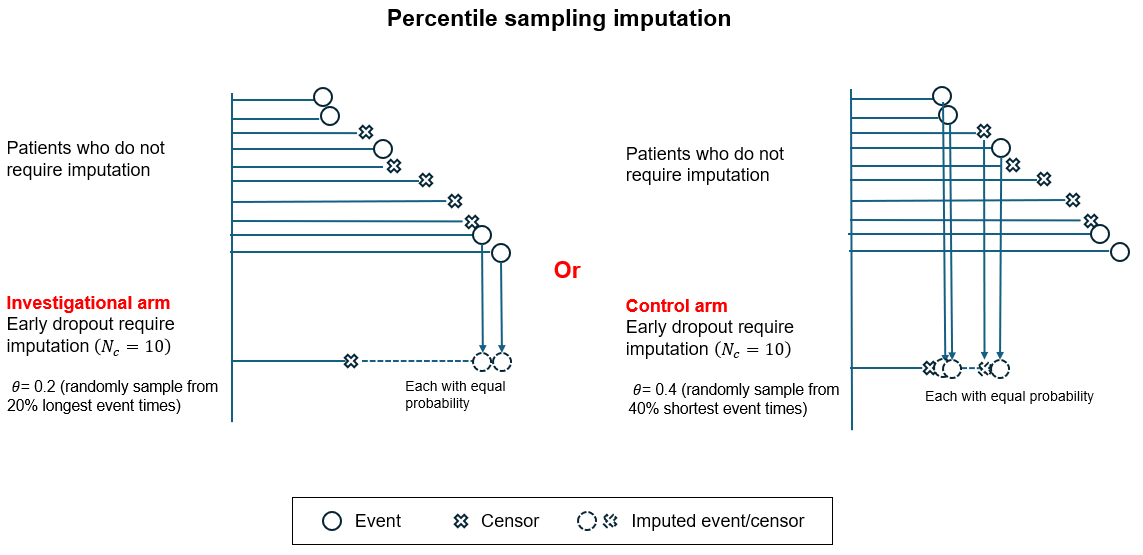}
    \end{subfigure}
    \caption{Illustration of ad hoc 1) landmark sampling and 2) percentile sampling imputations}
    \label{fig:illustration}
\end{figure}

\end{document}